# A micro-to-macroscale and multi-method investigation of human sweating dynamics

Cibin T. Jose,[1] Ankit Joshi,[1,2] Shri H. Viswanathan,[1] Sincere K. Nash,[3] Kambiz Sadeghi,[1,2] Stavros A. Kavouras,[4] and Konrad Rykaczewski[1,2*]

1. School for Engineering of Matter, Transport and Energy, Arizona State University, Tempe, USA
2. Julie Ann Wrigley Global Futures Laboratory, Arizona State University, Tempe, AZ 85287, USA
3. School of Life Sciences, Arizona State University, Tempe, AZ 85287, USA
4. Hydration Science Lab, College of Health Solutions, Arizona State University, Phoenix, AZ 85004, USA

*Corresponding author: konradr@asu.edu (Orcid ID: 0000-0002-5801-7177)

## Abstract

Sweat secretion and evaporation from the skin dictate the human ability to thermoregulate and thermal comfort in hot environments and impact skin interactions with cosmetics, textiles, and wearable electronics/sensors. However, sweating has mostly been investigated using macroscopic physiological methods, leaving micro-to-macroscale sweating dynamics unexplored. We explore these processes by employing a coupled micro-imaging and transport measurement approach used in engineering studies of phase change processes. Specifically, we employed a comprehensive set of "macroscale" physiological measurements (ventilated capsule sweat rate, galvanic skin conductance, and dielectric epidermis hydration) complemented by three microscale imaging techniques (visible light, midwave infrared, and optical coherence tomography imaging). Inspired by industrial jet cooling devices, we also explore an "air jet" (vs. cylindrical) capsule for measuring sweat rate. To enable near-simultaneous application of these methods, we studied forehead sweating dynamics of six supine subjects undergoing passive heating, cooling, and secondary heating. The relative dynamics of the physiological measurements agree with prior observations and can be explained using imaged microscale sweating dynamics. This comprehensive study provides new insights into the biophysical dynamics of sweating onset and following cyclic porewise, transition, and filmwise sweating modes, and highlights the roles of stratum corneum hydration, salt deposits, and microscale hair.

**Keywords:** sweat production, sweat evaporation, droplets dynamics, midwave infrared imaging, optical coherence tomography, galvanic skin conductance, stratum corneum hydration, salt deposits, ventilated capsule

**1. Introduction**

The characterization of sweating dynamics on length scales that span single sweat gland, duct, and pore to about two centimeters can provide insight into, and uncover potential routes to manipulate, the biophysical processes underlying human evaporative cooling [1,2], perception of wetness and thermal comfort [3], thermomechanical interactions with contacting objects [4–7] (from cosmetics [8,9] and allergens [10,11] to textiles [7,12] and wearable electronics/sensors [13,14]), and clinical diagnosis relying on detection sweating deficiency (e.g., for hypohidrosis, anhidrosis, and diabetes mellitus [15]). However, skin surface sweating dynamics are surprisingly understudied on the micro-to-macroscale, which falls in between studies focusing on sweat generation biochemistry and microfluidics of secretion [2,3,15–17] and those focusing on sweat rate variation across large skin regions [2,15,18–20] or whole-body level [21–25]. To address this gap, we employed coupled microscale imaging and macroscale transport measurement approach used in engineering studies of phase change processes [26,27] to investigate sweat transport, stratum corneum absorption, and skin surface dynamics. As shown in **Figure 1**, we employed a comprehensive set of "macroscale" physiological measurements (ventilated capsule for measuring sweat rate (SR), galvanic skin conductance (GSC), and dielectric epidermis hydration meter (HYD)) complemented by three microscale imaging techniques (macrophotography and midwave infrared (MWIR) skin surface imaging as well as optical coherence tomography (OCT) cross-sectional imaging). As we discuss below, these measurement techniques have been previously employed individually or in various combinations in sweat studies, while the imaging techniques have been mostly used individually to visualize mental sweating (i.e., short-lasting sweating due to non-thermal stimuli).

The production of sweat starts a few millimeters below the skin surface with the secretion of an isotonic liquid into the secretory coils of the sweat glands within the hypodermis [16,17], a process that can be measured using GSC [28–31]. The technique also detects the following movement of the liquid through the multiple parallel sweat ducts, since the presence of the weak electrolyte solution changes the effective electrical skin conductance measured over a few centimeters distance (the ion concentration typically decreases as it travels through the duct due to ion reabsorption [16]). Once near or on the skin surface, sweat can evaporate, accumulate, or in the initial stages of the process, be absorbed into the 10 to 20 $\mu$m thick outer skin layer, the stratum corneum, which is highly hygroscopic (can absorb up to 70% of water by volume [32,33]). In contrast to the deep probing depth of the GSC, the hydration meter reflects moisture content only within the outer ~50 $\mu$m of skin (effective properties averaged over a circle with about 1.5 cm diameter) [34], and has been previously applied along with GSC by Gerrett et al. [15] to distinguish between sweat transport through the duct and diffusion into the stratum corneum.

Gerrett et al. [15] combined GSC and HYD with visual observations of skin sweat accumulation and regional surface sweat rate measurements using absorbent pads with a 10-minute frequency. The absorbent pad method generally correlates well with cylindrical ventilated capsule sweat rate measurements (SR) [35]. However, we chose the latter for its high temporal resolution (~5-second dictated by humidity sensor [36]) and potential for near-surface sweat detection (vs. pads that only absorb sweat excreted onto the surface). The premise of the method is the evaporation of all sweat near or on the skin using a parallel flow of dry air, and it has been used to measure sweat rate for about a century [2,37,38]. Since cooling or drying using normally impinging air jets provides a multifold increase in heat and mass transfer rates as compared to parallel flow [39,40] and is extensively used in industrial cooling devices [41,42], we also developed a "air jet" (vs. traditional cylindrical) ventilated capsule and explored its use for sweat rate measurements. Instead of visual observations by researchers used by Gerrett et al. [15], we combined GSC, HYD, and SR measurements with three microscopic sweat imaging techniques.

Sweat emerging from pores is relatively difficult to see, so has been imaged using many non-visual approaches, including polymer imprinting [43,44], applying water-sensitive coatings on the skin [2,45,46] (or rolling films containing them across the forehead [47]), various permutations of photography (e.g., with a prism and angled flash) [2,48–50], midwave (MWIR: 3-5 $\mu$m) [36,51,52] and longwave (LWIR: 8-14 $\mu$m) [53–56] infrared microscopy, and optical coherence tomography (OCT) [57–61]. MWIR thermography has several advantages demonstrated in prior imaging of mental sweating (i.e., rapid and short-lasting in response to non-thermal stimuli) and in our recent work on thermal sweating on the forehead [36]. In particular, MWIR can detect even thin water films [36,62,63] because water has a peak in absorptivity at 3 $\mu$m [64]. We have demonstrated fast imaging (10 Hz or more) from onset to profuse sweating over a 2 cm$^2$ area of a passively heated subject's forehead with a resolution sufficient to capture single sweat pores (~20 $\mu$m) [36]. However, we have also observed that the MWIR contrast can diminish below utility levels once large sweat puddles and films are formed under a low evaporation rate, leading to a nearly isothermal surface. To address this issue, we cyclically alternate MWIR with OCT imaging (the two microscopes are too large to image the subject's forehead simultaneously). OCT provides cross-sectional images revealing sub-surface anatomical structures down to ~2-3 mm, including stratum corneum and the epidermal-dermal junction over various body regions [65–68] (it also provides macroscopic photographs of the scanned area, see **Figure 1a**). The technique has also been used to visualize sweat ducts, droplets, and thin water film applied to fingertips during mental sweating and hand grip exercises with resolution down to 10 $\mu$m [57–61]. However, the fingertip and other glabrous skin areas (e.g., palms and soles) have a special anatomical structure, including 0.5 mm thick stratum corneum and wide cone sweat pore openings that is different from non-glabrous skin, which is dominant in the thermal sweating and thermoregulation process. Therefore, the current study is the first to apply OCT to visualize thermal sweating on non-glabrous skin and across the entire sweating dynamics from onset to large sweat puddles during passive subject heating. Since salts present in sweat can alter sweat droplet evaporation dynamics on artificial surfaces [69] and deposit on the skin [10], we also study sweating dynamics during secondary subject heating following drying out of all sweat during a subject cooling stage.

Although there are prior examples of combining some of the methods, the current study provides a comprehensive combination of macroscopic measurement (GSC, HYD, SR) with microscale imaging techniques (MWIR, OCT, and macrophotography) applied to quantify microscale thermal sweating dynamics from onset to filmwise as well as drying and second sweating stages. For example, in the past Thomas and Korr [48] used GSC and photography (with prism and flash) to show that in early sweating stages, skin conductance increases linearly with the number of active sweat glands. Kato et al. [58] visualized sweat ducts and droplets using OCT and measured sweat rate using a ventilated capsule on side-by-side fingertips before and after handgrip exercises to show that the mean cross-sectional area of acrosyringia correlates with sweat rate and is higher in subjects with hyperhidrosis than in healthy controls. In our recent work, we used MWIR to image sweating through a sapphire window on a wind-tunnel-shaped ventilated capsule, which was designed not to evaporate all emerging sweat as traditional capsules but to control and measure the evaporation rate, and image related microscale dynamics, induced by more natural laminar airflow [36]. The MWIR imaging demonstrated how sweating progresses from cyclic dropwise mode [45,46,52,70–76], which occurs due to oscillatory sympathetic nerve activation of the sweat gland [76], to stratum corneum hydration driven transition to the filmwise mode [36]. In turn, combining evaporation rate measurements with the fraction of wet skin area obtained using analysis of over 25,000 MWIR images revealed that mass transfer coefficient in the dropwise mode is three times higher than in the filmwise mode, clearly highlighting the need for further investigations into microscale sweating and evaporation dynamics [36]. To this end, we investigate thermal sweating using GSC, HYD, SR, MWIR, OCT and macrophotography on foreheads of six supine subjects undergoing passive heating, cooling, and secondary heating (see Figure 1). This multi-pronged approach can highlight the respective utility of each method and reveal microscopic biophysical processes underlying changes in effective (i.e., area-averaged) macroscopic property measurements.

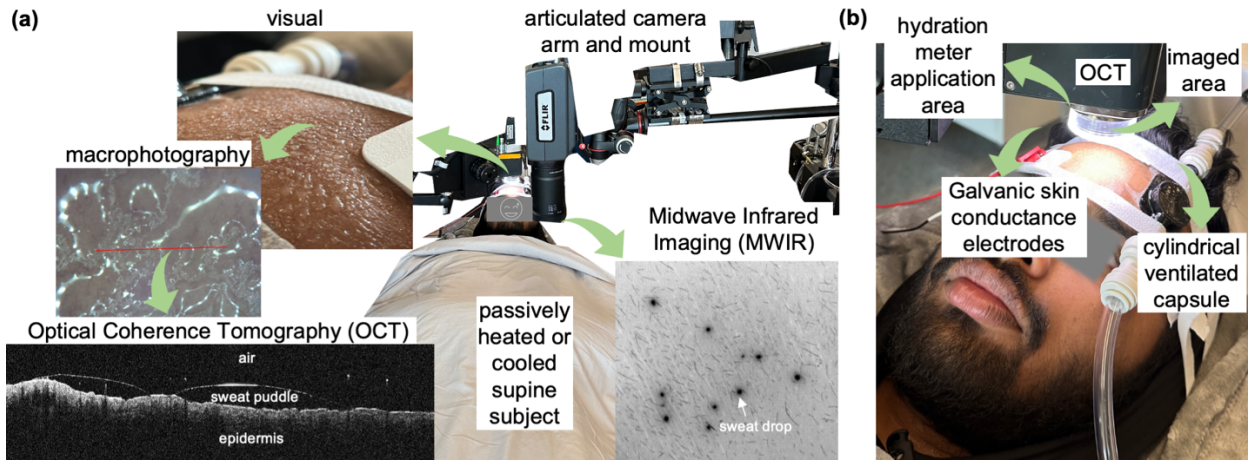

**Figure 1.** Image showing **(a)** overview of the experimental setup including imaging methods (MWIR and OCT/macrophotography lenses are cyclically placed above the imaged area) and (b) of instrumented subject forehead during the first sweating stage.

## 2. Methods

### 2.1 Overview

We developed a custom experimental setup to enable simultaneous measurements using GSC, HYD, and SR and periodically altered MWIR and OCT/macrophotography (see **Figure 1a**). The MWIR and OCT have relatively large lenses, so they could not be simultaneously positioned above the skin area of interest. In addition, both have a narrow depth of field (below 1 mm), therefore requiring some focus adjustment at the start of every imaging segment (every ~3-5 minutes). To accommodate the imaging requirements, periodic imaging method switching, and room for GSC, HYD, and SR, we conducted the experiments on supine subjects' foreheads. **Figure 1a** shows that we placed a reclining chair (Paddie Electric Height Adjustable Bed Chair, Electric Lift Massage Table 3-Section Folding) in between two laboratory benches onto which we mounted two swinging arms (AmScope Articulating Stand with Clamp and Focusing Rack for Stereo Microscopes with modified attachments) to which MWIR and OCT lenses were attached. The reclining chair setup and subject passive heating/cooling method are described in depth in prior work [36] and were based on procedures by Meade et al. [37] and Rutherford et al.[38].  In particular, heated blankets were placed below and above the subject, who was wearing a water perfused full body garment (Compcooler Full Body Cooling Garment with stretchable fabric XS/S and M/L, hand washed and dried in between each experiment), through which either hot (47°C) or cold (25°C) water was passed from separate constant temperature baths (VWR AP28R-30-V11B). The subject was separated from the blankets using 76.2 cm wide waterproof exam paper (TOA Disposable Polypaper waterproof exam paper) to prevent their saturation with sweat. While the central part of the forehead was saved for imaging, the GSC electrodes and ventilated capsule were attached to the sides, as illustrated in **Figure 1b**. Below, we describe the cylindrical and jet capsule design, while the details of the experimental protocol, study participant, GSC, HYD, SR, MWIR, and OCT/macrophotography are covered in the **Supplemental Material**—the **SM**.

### 2.2 Sweat rate measurements using ventilated capsules

As in our prior work [36], we digitally designed the cylindrical and air jet capsule geometries (see **Figure 2a-b**) using Solidworks 2024 software and 3D printed them (Prusa i3 MK3S with 0.1 mm 'detail' setting, 30% infill) using polyethylene terephthalate glycol (PETG, Overture 1.75 mm) filament. To seal potential gaps, we coated all printed parts three times with acetone-diluted epoxy (XTC 3D) and cured the parts for 24 hours. Following the cylindrical capsule designs used in physiological studies for over a century [2,18,38], the dry inlet air enters and leaves the capsule on opposite sides of the cylindrical enclosure (see **Figure 2a**). We made two sizes of cylindrical capsules, one with final 1.7 cm$^2$ and one with 2.7 cm$^2$ open areas. The capsule geometry included an inlet and outlet with inner (ID) and outer (OD) diameters of 5.50 and 9.50 mm, which were connected to 6.35 mm inner (ID) and 9.5 mm outer (OD) diameter tubing using 9.5mm quick-connectors (LASCO, Push on fittings) (see **Figure 2a** and mechanical drawing in the **SM**).

The air jet capsule had a more complex air flow path (see **Figure 2b** and the **SM**), with inlet air entering a large plenum above the slotted acrylic to enable uniform flow across the 15 mm long slots. The air accelerates through the slots to impinge normally onto wet surface at high velocity. We used a laser cutter (Full spectrum laser PS20) to cut five slots (outlet width) within a 1 mm thin acrylic sheet, which was inserted into the side hole in the jet capsule and sealed using the epoxy. We designed the slot width (W=0.77 mm), slot-to-slot separation distance (S=2 mm), and slot outlet to skin surface height (H=2 mm) by balancing our fabrication realities with broad recommendations for optimal S/H and H/W ratios (S/H=1.4 and H/W=10) that maximize the heat/mass transfer coefficients [39,40]. These dimensions resulted in final S/H=1, S/W=2.6, and H/W=2.6 ratios.

After impinging normally on the skin, the humid flow was parallel to the skin surface and transitioned through a 1.5 mm high connecting chamber into the larger outlet. To enable a compact size while still ensuring that the humidity probe is not impacted by either distance from the evaporation area or from the outlet to the room [36], the outlet section was curved by 180º ("U-bend"). We also made air jet capsules with the same evaporation area to probe and probe to outlet distances without the U-bend.

All capsules were equipped with a compact humidity probe (Vaisala HMP9) with high accuracy (±0.8% relative humidity or ±0.18 g·m⁻³) with a digital transmitter (Vaisala Indigo202). We inserted the probe into a 7.9 mm hole on the side of the evaporation section using a grommet (Buna-N, MS 35489-91). We used Ultra Zero Air (Airgas, less than 2 ppm of water) from a high-pressure cylinder. We regulated the flow rate using a digital mass flow controller (Alicat Scientific, MC-1SLP-D-DB9M/5m, standard accuracy of ±0.1% of full scale or ±0.6% of the reading) with a typical airflow rate of 0.5 L·min⁻¹ [18,37,38]. Prior to human trials, we quantified the evaporative performance of the various ventilated capsules using the artificial sweating surface described by Jaiswal et al. [36], which had a 1 cm² square water-thin film evaporating from a laser-etched acrylic that was heated from below. For human trials, the capsules were attached to and sealed with the skin using double-sided skin tape (skin-compatible double-sided tape, BearKig). The tape adhered well to most subjects' skin but required changing between the first and second heating stages for a few subjects.

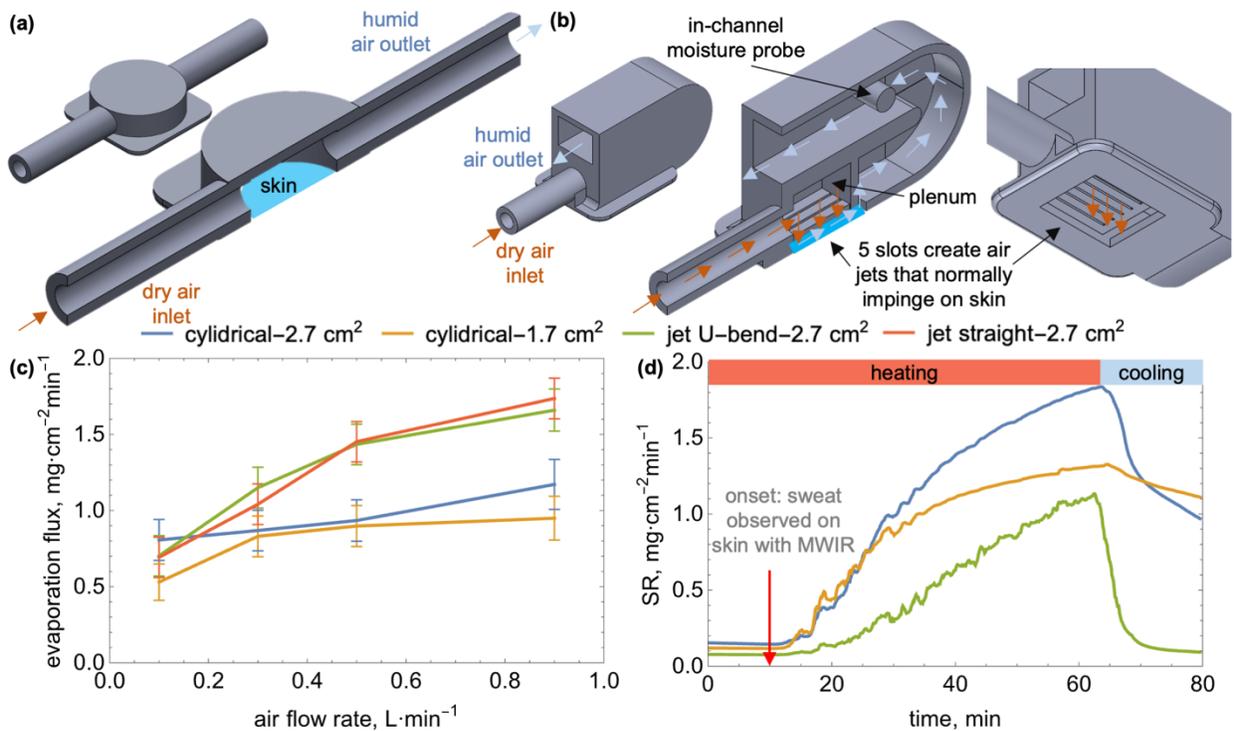

**Figure 2. (a)-(b)** whole and cross-sectioned three-dimensional models of **(a)** the cylindrical and **(b)** air jet capsules; **(c)** the evaporation flux as a function of dry air flow rate measured for different ventilated capsules placed on artificial sweating surface (1 cm² square and heated water thin film at 34ºC), and **(d)** sweat rates (SR) for subject 7 comparing two cylindrical and U-bend air jet capsules with air flow rate of 0.5 L·min⁻¹.

## 3. Results

### 3.1 Sweat rate measurement using cylindrical and air jet ventilated capsules

Before implementing the complete array of methods, we compared the evaporation rates attained using the cylindrical and air jet capsules on an artificial sweating surface and in a pilot human trial. As shown in the plot in **Figure 2c**, the evaporation fluxes measured using the artificial sweating surface for the air jet capsules were generally about 50 to 75% greater than those for the cylindrical capsules (e.g., 0.8 mg·cm$^{-2}$min$^{-1}$ for cylindrical vs. 1.4 mg·cm$^{-2}$min$^{-1}$ for jet capsules at an air flow rate of 0.5 L·min$^{-1}$). In contrast to the significant impact of the capsule type, the shape of the air jet capsule outlet (straight vs. U-bend) and the size of the evaporation area for the cylindrical capsules had a minor impact on the evaporation rate. However, the capsules' relative performance was reversed during the human trials.

The plot in **Figure 2d** shows that while the three capsules measured similar imperceivable perspiration flux (i.e., via diffusion through skin that is typically below 0.15 mg·cm$^{-2}$min$^{-1}$ [18]), past onset of sweating detected through MWIR imaging (with onset time adjusted to 10 minutes in the plot to facilitate interpretation), the sweat rate (SR) measured using the air jet capsule is substantially smaller than those measured using the cylindrical capsules. For example, after 20 minutes of sweating (i.e., at 30 minutes on the plot), the air jet capsule measured SR of only 0.3 mg·cm$^{-2}$min$^{-1}$, while the two cylindrical capsules measured SR of 0.9 to 1 mg·cm$^{-2}$min$^{-1}$. Afterward, the SRs measured using the two cylindrical capsules diverged, with SR measured using the smaller capsule slowly saturating, while the SR measured using the larger capsule continued to increase. The peak SRs measured using 1.7 cm$^2$ cylindrical, 2.7 cm$^2$ cylindrical, and 2.7 cm$^2$ air jet capsules after 55 minutes of sweating were 1.1 mg·cm$^{-2}$min$^{-1}$, 1.3 mg·cm$^{-2}$min$^{-1}$, and 1.8 mg·cm$^{-2}$min$^{-1}$. Past that point, the subject was cooled, resulting in a decrease in the sweat rate, which also varied by capsule type and size. Within 5 minutes, the SR measured using the air jet capsule decreased to the pre-onset baseline. Within 15 minutes, the SR measured using the small and large cylindrical capsules decreased from 1.3 to 1.1 mg·cm$^{-2}$min$^{-1}$ and from 1.8 to 1 mg·cm$^{-2}$min$^{-1}$, respectively. Putting the sweating and drying results together demonstrates that the air jet capsule is very effective at evaporating sweat and cooling the skin surface (the air is at lab temperature of 22ºC). Since the flow rates and exposed skin areas are matched (between the 2.7 cm$^2$ capsules), the jet capsule likely locally over-cools the surface, resulting in delayed and lower SR during subject heating owning to lower sweating drive and water saturation pressure at the skin, and rapid SR decrease during subject cooling. Considering that the smaller cylindrical capsule appears to have saturated, we conducted the rest of the human trails employing the large cylindrical capsule for sweat rate measurements.

### 3.2 Representative sweating dynamics across the heating-cooling-heating stages

**Figures 3** and **4** demonstrate that the SR, HYD, and GSC measurements across heating-cooling-heating stages generally follow the same trend, which can be clearly correlated to microscopic sweating dynamics and regimes observed through imaging. In particular, the three macroscopic measurements all begin to increase from baseline in a 1-to-3-minute period surrounding the onset of sweat emerging out of the pores (after 14 minutes from the start of heating in the case of subject 2, shown in **Figures 3** and **4**), as observed through MWIR. The GSC begins to increase first, about 1.2 minutes before the surface sweating onset. In contrast, the two other measurements start to increase after the onset, with SR increasing from 0.18 mg·cm$^{-2}$min$^{-1}$ baseline after 0.85 minutes and HYD increasing from the 60 to 70 AU baseline after 4.5 minutes (note that the HYD was measured every 2 to 2.5 minutes). For the next 15 minutes, the SR and HYD increase at a nearly constant rate (about 0.5 mg·cm$^{-2}$min$^{-1}$ and 50 AU per 10 minutes). In contrast, past the initial rise of about 3 μS, the change in GSC was very gradual. This trend is reversed, starting 15-to-20 minutes from the onset and continuing to the peak that occurs when the subject is cooled (about 40 minutes from onset). In particular, in this period, the SR and HYD start to saturate while the GSC increases rapidly and substantially (about 12 μS in 5 minutes). These macroscopic measurement trends clearly correspond to the microscopic sweating dynamics observed through imaging.

The first 15 minutes past onset corresponds to the mode previously referred to as cyclic dropwise [36], in which sweat periodically emerges and evaporates from a gradually increasing number of pores but does not spread beyond them. This behavior is evident in the high-contrast circles in MWIR images in **Figure 3b** and **Movie 1**, but is unperceivable using macrophotography and challenging to observe using OCT. The cross-sectional images in **Figure 3c** show that in this period, about 50 to 150 μm stretches of the liquid-air interface can be distinguished from the stratum corneum using the OCT. Morphologically, these features are not the often-imagined hemispherical droplets. Instead, sweat is visible in OCT as conical features with shallow convex menisci filling small skin crevices in which the pores are located. After about 15 minutes from onset, the cyclic behavior dampens with sweat spreading beyond the pores and gradually transitioning into the filmwise mode. Considering these results, we will refer to dropwise mode as the porewise mode since the latter better captures the liquid sweat being confined to near the pore opening in this mode. In the MWIR images, the transition mode between the porewise and filmwise modes is exhibited by a gradual increase and occasional merging of the original sweat areas along with a decrease in the contrast between the wet (cooler) and dry (warmer) areas. As revealed remarkably clearly by the OCT, these features are again not near hemispherical droplets but shallow (contact angle of 20 to 40º and maximum depth of around 0.5 to 1 mm) and irregularly shaped large puddles that can easily stretch more than 5 mm. Eventually, in the filmwise stage, the sweat puddles mostly merge and form a film whose exterior is nearly at the same temperature as skin, which diminishes the MWIR contrast and utility of infrared imaging. The patches of sweat film are much thinner (10 to 200 μm) and essentially fill or cover underlying skin texture and features. During the transition and filmwise stage, sweat can be observed through macrophotography, at first through local increase in reflectivity and later with clearly visible sweat puddles.

Once the subject is cooled, the liquid sweat evaporates in a filmwise manner, being unobservable with any of the imaging techniques after 5 to 10 minutes (see **example in the SM**), which also corresponds to substantial decrease in all the macroscopic measurements. However, while GSC proceeds to plateau within 5 minutes at a value substantially higher than the pre-onset baseline, the HYD and SR continue to decrease. The HYD takes about 10 to 15 minutes past the peak to return to its pre-onset baseline, while the SR takes about 40 minutes. In addition, the SR shows two subsequent stages of drying, each involving a fast decay followed by a slower linear decrease. We began the second subject heating stage after all the macroscopic measurements reached a plateau or baseline.

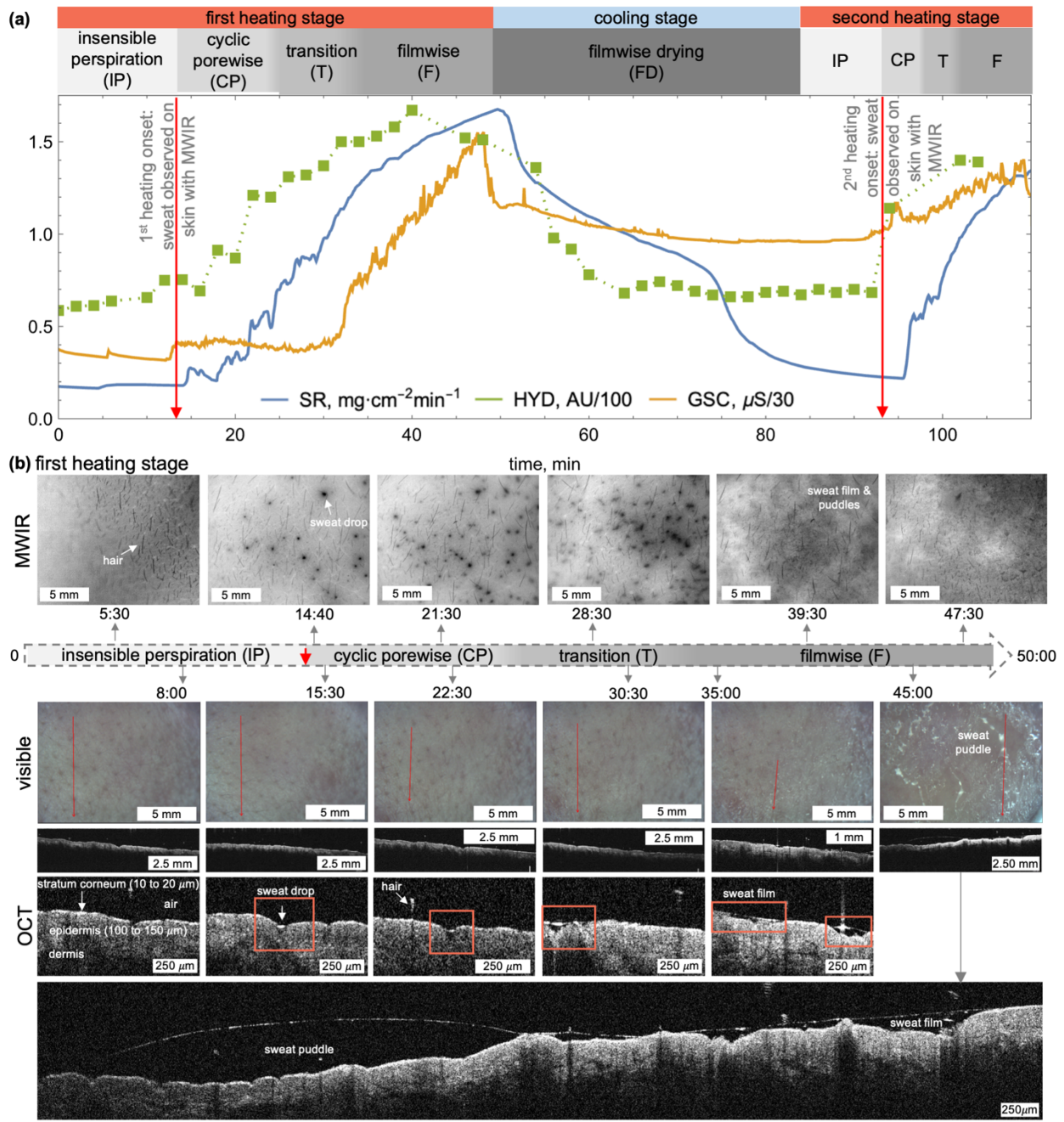

**Figure 3.** Representative sweating and drying dynamics during subject 2 heating, cooling, and second heating stages: **(a)** the sweat rate (SR), galvanic skin conductance (GSC), dielectric epidermis hydration (HYD) plotted against time along with annotations about microscopic sweating modes observed through imaging (note: the GSC and HYD measurements are scaled as indicated in legend to facilitate comparison); the corresponding **(b)** MWIR and **(c)** macrophotography and OCT images for first heating stage (the redline in photograph indicates the OCT scan).

The sweating dynamics during the second subject heating generally followed the same trends as during the first stage. However, they occurred more rapidly and with a significant difference in the liquid sweat morphological evolution during the transition stage. In particular, **Figure 3a** shows that the second onset in surface sweating occurred within just 8 minutes of the cooling-to-heating switch (vs. 14 minutes in the first heating cycle), with GSC again increasing about 2.1 minutes before pore activity was detected using MWIR. As in the first cycle, the HYD and SR increase shortly after the onset (HYD after 1.2 minutes and SR after 2.0 minutes). However, compared to the first onset, the HYD and SR increase much more rapidly, reaching slightly below their prior peak values within 10 to 15 minutes of the second sweating onset. The GSC follows the same general trend as in the first heating stage but with an offset baseline and condensed timeline. Specifically, the GSC nearly reaches a plateau after about 3 to 4 µS initial increase. However, instead of taking about 15 minutes as in the first heating stage, during the second subject heating stage, the GSC plateau only lasts about 5 minutes. Subsequently, the GSC increases, reaching its prior peak value. These macroscopic measurement trends can be readily explained using microscopic imaging. The images in **Figure 4** show that the cycling porewise mode occurs only for 5 minutes past the second sweating onset. Subsequently, sweat emerges past the pores, gradually transitioning from cyclic porewise to filmwise mode. In contrast to the first heating stage, the liquid sweat does not accumulate in large puddles but forms gradually increasing film patches, which merge to cover most of the surface within just about 10 minutes after the onset. Next, we discuss the measurement and imaging trends observed across all subjects.

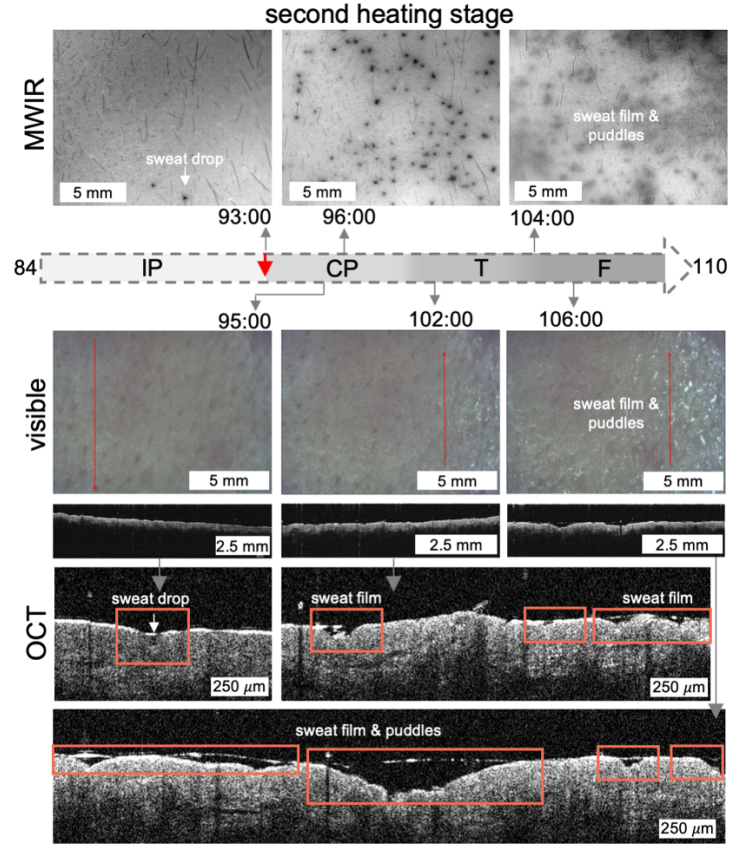

**Figure 4.** The MWIR, macrophotography, and OCT images corresponding to the second heating stage of the trial shown in Figure 3 (the red line in the photograph indicates the OCT scan).

### 3.3 Sweating dynamics for all subjects

The trends during the heating-cooling-second heating stages covered above for subject 2 were generally observed for almost all subjects (see **Figure 5** and **Table 1** for SR, HYD, GSC compiles for all subjects and the **SM** for individual subject measurements and imaging). The baseline SR values were in the typical 0.1 to 0.2 mg·cm$^{-2}$min$^{-1}$ range before the first heating stage and were mildly elevated to 0.2 to 0.25 mg·cm$^{-2}$min$^{-1}$ range before the second heating stage. The baseline HYD values varied between 30 and 70 AU before both sweating stages. The variation in the baseline GSC was much greater, therefore following Gerrett et al.[15], we reported values as an increase from the pre-heating baseline for all subjects (i.e., ΔGSC). The ΔGSC begins to increase above the baseline signal first, on average 1.7±0.8 minutes (±1 standard deviation) before sweat is detected on the skin using MWIR. A measurable increase in SR and HYD occurred on average 1.4±0.7 minutes and 2.7±1.8 minutes after the observation of sweat on the skin, correlating with the increasing number of active pores displaying the cyclic porewise behavior.

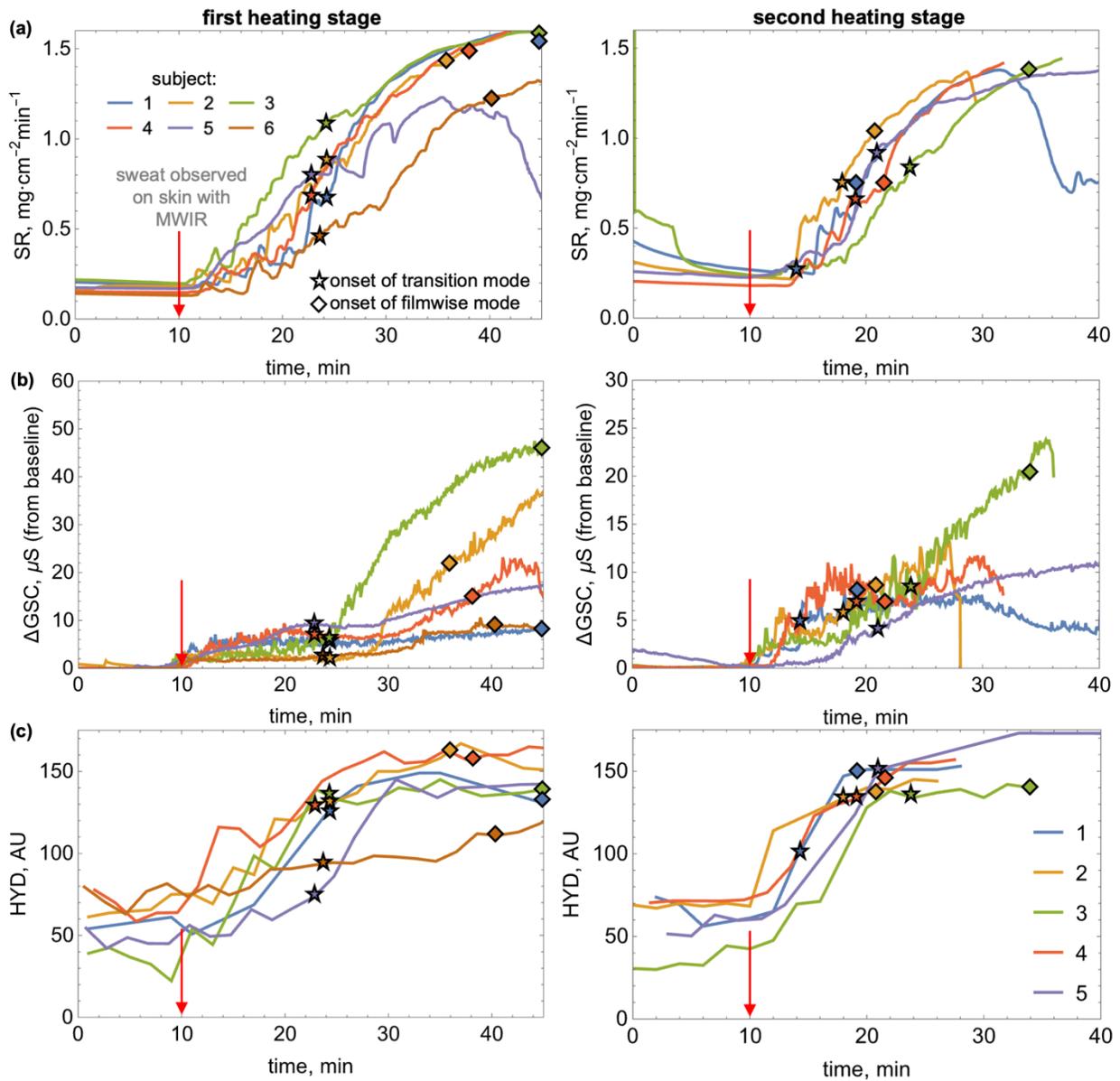

**Figure 5. (a)** the sweat rate (SR), **(b)** the galvanic skin conductance (ΔGSC), **(c)** dielectric epidermis hydration (HYD) plotted against time along with annotations about microscopic sweating modes observed through imaging (note: the baselines for ΔGSC for first and second heating stages are different, see Figure 3) during first and second heating stage measurements for all subjects showing. The timeframe is adjusted so that the onset of sweat detected with MWIR is at 10 minutes (note: the GSC and HYD measurements are scaled as indicated in the legend to facilitate comparison).

During the first heating stage, the cyclic porewise mode occurred for 13 to 15 minutes after onset for all subjects, which was followed by 10 to 20 minutes of the transition mode before shifting to the filmwise mode. At the start of the transition mode the SR varied substantially from subject-to-subject with a range of 0.5 to 1 mg·cm$^{-2}$min$^{-1}$, which increased to 1.2 to 1.5 mg·cm$^{-2}$min$^{-1}$ during shift to the filmwise mode. In contrast to the relatively large scatter in absolute SR values, the difference between SR at the shift to transition and to filmwise modes was much more consistent across all individuals, varying from 0.5 to 0.8 mg·cm$^{-2}$min$^{-1}$. Post the initial rise in the range of 1.5 to 9 µS, the ΔGSC changed very gradually until it increased rapidly during the transition mode, reaching in a few cases over 20 to 40 µS. For most subjects, the start of the transition mode correlated to being near or above 80 to 120 AU, with only a mild increase observed for the rest of the experiment. In general, only the sweating dynamics of subject 5 deviated from this trend. In particular, despite reaching a high SR and HYD, this subject never fully transitioned into the filmwise mode with clearly distinguishable individual sweat accumulated around the pores (see the SM). Once subject cooling was initiated, sweat film detectable using MWIR or OCT evaporated within 5 to 10 minutes, which was followed by a more extended SR decrease period. The second heating stage was started once SR and HYD reached their baselines.

During the second heating stage, the macroscopic measurement trends (SR, ΔGSC, and HYD) and the start of their increase compared to the MWIR-detected sweating onset were comparable to the first heating stage. However, the timing of the start of the transition and filmwise modes were more scattered. In all, the sweating occurred faster after the start of subject heating and switched from cyclic porewise to transition modes and later to filmwise mode quicker. In particular, the first switch, from cyclic porewise to transition modes, occurred only within 4 to 14 minutes after onset (vs. 14 minutes for nearly all subjects in the first stage) and was followed by the second switch, from transition to filmwise modes, within only 3 to 11 minutes (vs. 10 to 20 minutes in the first stage). The SR and HYD also increased faster than in the first stage. On average, within 10 minutes of starting to increase (about 2 minutes after MWIR onset), the SR increased, on average, by 0.76 mg·cm$^{-2}$min$^{-1}$ compared to only 0.47 mg·cm$^{-2}$min$^{-1}$ for the equivalent period during the first heating stage. Similarly, within the same period, the HYD increased on average by about 90 AU compared to only 65 AU during the first heating stage. From there, the SR and HYD reach about the same peak values as during the first heating stage. The GSC displayed an initial rise in the range of 1.5 to 9 µS above the baseline, also seen during the first heating stage, followed by a more gradual increase. In contrast to the first stage, however, only subject 3 had a drastic increase from that range (vs. most during the first stage). This behavior can likely be explained when looking at the absolute GSC values. In particular, the starting baselines for the GSC were much higher during the second heating stages (see example in **Figure 3**), which slightly obstructs that the absolute GSC reaches about the same peak values during both stages (see the **SM**). As for the representative results for subject 2, the only major difference between first and second-stage microscale sweating dynamics for the rest of the subjects was that during the transition mode, sweat spread beyond the pores in the form of a low-contact angle film rather than slightly higher contact angle puddles. As during the first stage, only subject 5 never reached the filmwise stage. Next, we briefly cover additional observations from microscale imaging.

**Table 1.** The summary of microscopic sweating mode onset times and corresponding SR, GSC, and HYD measurements observed during the first and second heating stages of subjects 1 to 6. F-filmwise, T-transition, CP-cyclic porewise modes. The timeframe is adjusted so that the onset of sweat detected with MWIR is at 10 minutes for both stages.

| | first heating stage | | | | | | second heating stage | | | | |
|---|---|---|---|---|---|---|---|---|---|---|---|
| subject | 1 | 2 | 3 | 4 | 5 | 6 | 1 | 2 | 3 | 4 | 5 |
| CP mode onset: ΔGSC, min | 7 | 8.8 | 7 | 7.5 | 7.5 | 9 | 8.5 | 7.9 | 9.2 | 8.7 | 10 |
| CP mode onset: SR, min | 11.7 | 10.8 | 10.7 | 11 | 10.5 | 11 | 12 | 12 | 12 | 13.2 | 10.5 |
| CP mode onset: HYD, min | 16 | 14.5 | 11 | 11.5 | 10.7 | 18 | 12 | 11.2 | 12 | 11.5 | 13 |
| CP mode onset: MWIR, min | 10 | 10 | 10 | 10 | 10 | 10 | 10 | 10 | 10 | 10 | 10 |
| ΔGSC, µS | 0 | 0 | 0 | 0 | 0 | 0 | 0 | 0 | 0 | 0 | 0 |
| SR, mg·cm$^{-2}$min$^{-1}$ | 0.18 | 0.19 | 0.22 | 0.14 | 0.17 | 0.13 | 0.26 | 0.23 | 0.24 | 0.18 | 0.22 |
| HYD, AU | 50 | 65 | 22 | 63 | 56 | 76 | 61 | 68 | 45 | 72 | 59 |
| T mode onset, min | 25 | 25 | 25 | 23 | 23 | 24 | 14 | 18 | 24 | 19 | 21 |
| ΔGSC, µS | 5.3 | 1.6 | 4.65 | 7.1 | 9.1 | 1.7 | 1.75 | 5.7 | 9.9 | 8.25 | 3.9 |
| SR, mg·cm$^{-2}$min$^{-1}$ | 0.76 | 0.87 | 1.1 | 0.718 | 0.79 | 0.49 | 0.26 | 0.77 | 0.88 | 0.65 | 0.89 |
| HYD, AU | 130 | 132 | 117 | 139 | 82 | 95 | 81 | 133 | 134 | 133 | 150 |
| F mode onset, min | 45 | 35 | 45 | 38 | N/A | 40 | 19 | 21 | 35 | 26 | N/A |
| ΔGSC, µS | 3.9 | 19.9 | 46 | 14 | | 8.7 | 6.5 | 8.5 | 23 | 8.9 | |
| SR, mg·cm$^{-2}$min$^{-1}$ | 1.62 | 1.05 | 1.6 | 1.5 | | 1.21 | 0.6 | 1.0 | 1.4 | 1.19 | |
| HYD, AU | 149 | 145 | 139 | 157 | | 112 | 153 | 142 | 140 | 155 | |

*3.4 Microscale hair and sweating*

Although often not visible, the skin on the forehead is covered with microscopic hairs, which interact with sweat emerging on the skin in several ways. The MWIR sequence of images in **Figure 6a** shows several hairs within puddles of sweat during the late transition mode, with some standing up while others stick to the skin surface. The sequence illustrates how two hairs are forced to bridge when two puddles coalesce. Using OCT, we often imaged hair sticking to the skin surface during the transition and filmwise modes (see two examples in **Figure 6b**). In addition to interactions of sweat with hairs during these later stages, we also occasionally observed sweat wicking onto hair during the cyclic porewise stages. For example, the MWIR image sequence in **Figure 6c** clearly shows sweat wicking from a pore at the hair base onto its shaft and evaporating (or receding from it) during a single about 30-second cycle.

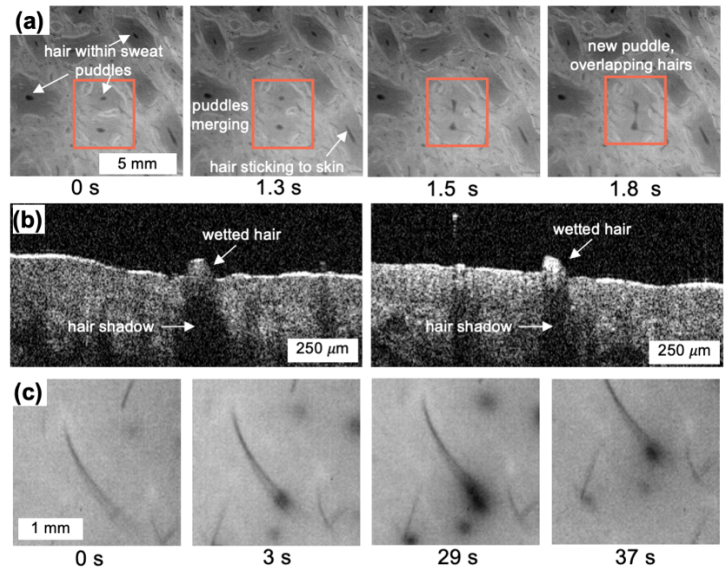

**Figure 6. (a)** MWIR image sequence showing multiple hairs wetted by sweating during late transition mode, including highlighted merging puddle-induced bridging of two hairs, **(b)** typical OCT image of a wetted hair sticking to skin, and **(c)** MWIR image showing sweat wicking onto and evaporating from a hair during the cyclic porewise stage.

## 4. Discussion

Following Gerrett et al. [15], we plot the macroscopic measurements against each other (i.e., eliminate the time component), also indicating our microscopic imaging observations. The SR vs. ΔGSC, SR vs. HYD, and ΔGSC vs. HYD plots in **Figure 7a-c** show that our measurements in the first and second heating stages mostly agree with those of Gerrett et al. [15], but there are noticeable differences between both macroscopic trends and microscopic onsets between the two heating stages. As Gerrett et al. [15], we did not observe any major differences between male and female subjects, with the only outlier from the general trend being the lack of mode shift from transition to filmwise for subject 5. The most evident difference between prior literature and our work relates to the shift of the increase in ΔGSC in the ΔGSC vs. HYD plot by about 25 AU in our work (see **Figure 7c**). This difference stems from a minor difference in the hydration measurement protocols. In particular, once sweat was visible on the skin, Gerrett et al. [15] wiped the area prior to the HYD measurement, while we did not, resulting in slightly higher measurement values that include a contribution of a water film. However, since in all the relevant cases, the HYD values are substantially above 100 AU, which implies a fully saturated stratum corneum on this arbitrary scale, the shift between our and Gerrett et al. [15] ΔGSC vs. HYD results is inconsequential. In contrast, the differences in macroscopic measurements and imaging observations between the two heating stages have implications on the interpretation of stratum corneum hydration dynamics and the key role of salt deposited from evaporated sweat, which we will discuss next.

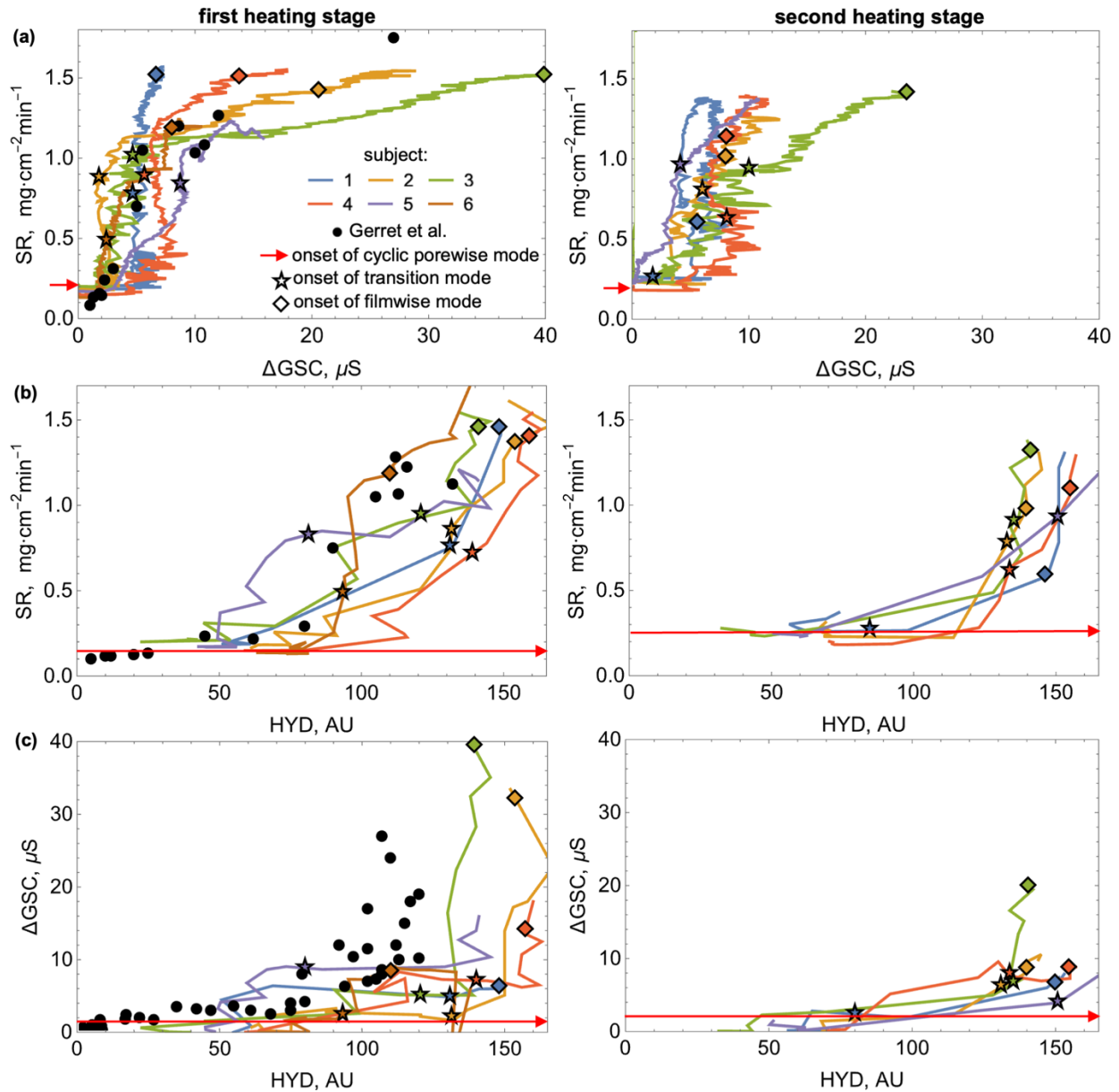

**Figure 7.** The relative variation of macroscopic measurements with the three sweating modes onsets indicated plotted along with corresponding data from Gerrett et al. [15] **(a)** the sweat rate (SR) vs. the galvanic skin conductance (ΔGSC), **(b)** SR vs. dielectric epidermis hydration (HYD), and **(c)** HYD vs. ΔGSC.

### 4.1 Onset of sweating and stratum corneum hydration

While prior studies have postulated [77] and concluded based on a comparison of HYD, GSC, and visual observations [15] that stratum corneum begins to hydrate from underneath prior to sweat secretion onto the skin, our observations provide a more nuanced view. In particular, Gerrett et al. observed a gradual increase in ΔGSC vs. HYD curve (from 0 to about 5 μS with HYD increasing up to about 50 to 75 AU in **Figure 7c**), which occurred before sweat was visibly noticeable (i.e., by eye). However, with the MWIR, which is multifold times more sensitive to sweat than visual inspection, we repeatedly observed sweat within pores about 1 to 2 minutes after an increase in the GSC signal and a few minutes before any meaningful increase in the HYD measurements. After onset, sweating occurs in the cyclic porewise mode, with the number of active pores increasing over time, which is also correlated to the increase in HYD. This measurement provides value averaged over a large area. Therefore, the observed HYD increase did not correspond to a homogenous stratum corneum hydration from within the body but was instead due to the highly localized and cyclic presence of sweat at the pores, which progressively hydrates the surrounding stratum corneum, eventually leading to the emergence of sweat beyond its pore vicinity. In contrast to the common "picture" of a hemispherical sweat droplet above the pore [16,17,36,52,55,69], OCT demonstrated that in the porewise mode, sweat has nearly "flat" or slightly concave meniscus when cyclically reaching the top of the duct (see **Figure 2** and the SM). As illustrated in **Figure 8a**, it is during these periods that the sweat is in contact with and hydrates the stratum corneum.

### 4.2 Shifts from cyclic porewise to transition to filmwise modes and role of deposited salts on the skin

The major differences between shifts from cyclic porewise mode to transition mode and later to filmwise mode and the corresponding sweat morphologies during the first and second heating stages illuminate the substantial role of salts deposited from evaporated sweat. Specifically, during the first heating stage, the cyclic porewise mode lasted about 15 minutes for all subjects, which was followed in about 10 to 15 minutes by the transition to the filmwise mode. Even when the temporal variation is removed, the shifts in the modes are reasonably well clustered for the first heating stage in the SR vs. ΔGSC and SR vs. HYD plots in **Figure 7**. As we illustrated in **Figure 8a**, the periodic sweat contact with the stratum corneum during the 14 minute cyclic porewise mode likely corresponds to the time required to hydrate this outer layer skin near the pore fully. It takes about 5 minutes of constant contact with water for the stratum corneum to fully hydrate [33], therefore, the on-and-off contact extends the time required for complete hydration in the area surrounding the pore to 14 minutes. The associated increase in surface wetting facilitates the spreading of sweat outside the pore, initiating the shift to the transition mode.

When the sweat rate is higher than its evaporation rate, sweat accumulates during the transition mode into shallow puddles with thicknesses of 0.1 to 1 mm and contact angles of 20° to 40° that progressively spread over the surface. The horizontal spreading rate of the puddles is controlled by the lateral rate of sweat diffusion within the stratum corneum and its hydration from the puddle's edge (see **Figure 8b**). As in our prior observations [36], the process can also proceed within crevices in the skin to bridge multiple pores. As illustrated by the "hairy" example in **Figure 6a**, an increase in sweat surface coverage can also occur through the coalescence of two large puddles. Eventually, the puddles cover most of the skin surface and often transition into a film (see **Figure 8d)** with a low contact angle that is very thin (less than 0.1 mm), besides regions in which it fills in any deeper skin surface topology (e.g., the large crevice in the center of the OCT image in the bottom of **Figure 2b**). In some subjects, the prevalence of large puddles persists longer than in others, leading to a delayed transition to the filmwise mode, despite covering most of the skin surface (see, for example, data for subject 1 in **Figures 5** and **7** and corresponding images in the **SM**). It is important to note that such a delayed transition might be partially an artifact of our experimental setup. Puddles that can accumulate on the horizontally oriented forehead of supine subjects are likely to shed with gravity in other postures or due to motion during various activities. As many have experienced in daily activities, shedding puddles leave behind a film "trail" of sweat. Unfortunately, the current limitations of our experimental setup do not allow us to investigate the shedding aspect. In either case, whether the subject's forehead is covered by a film or a mixture of a film and puddles, the sweat evaporates quickly after the cooling of the subject is initiated, leaving behind predominantly salt deposits [16] (see **Figure 8d**) that are likely one of the key factors underlying substantially different sweating dynamics during the second heating stage.

In the second heating stage, not only does the onset of sweating occur quicker and its rate increases faster than in the first heating stage, but most likely due to the presence of salts on the skin, the sweat emergence out of the pores onto the skin in a film rather than as shallow puddles observed in the first heating stage. Since the sweat rate is a function of core and skin temperature [24], faster sweating is likely induced by the already elevated core temperature (the heating stage is longer and provides a higher heat transfer rate than the cooling stage). Despite the generally faster sweating in the second heating stage, the temporal (**Figure 5**) and the respective (**Figure 7**) changes in the SR, ΔGSC, and HYD mostly overlap between the two heating stages. A notable difference is that the changes in HYD during the second stage for all subjects are much less scattered than during the first stage (see **Figure 5c**), even when compared for particular sweat rates (see **Figure 7b**). In contrast, the shifts between the sweating modes are highly dispersed across the subjects and in time. For example, the cyclic porewise to transition mode shift occurs between 4 to 14 minutes after onset for various subjects rather than after 14 minutes for nearly all subjects in the first stage. The following shift between transition and filmwise modes is also highly dispersed, occurring within the next 3 to 11 minutes (vs. 10 to 20 minutes in the first stage). This high variability in mode shifts between subjects is preserved even if the temporal dependence is removed, as seen in **Figure 7**. These trends can likely be explained through sweat interactions with the salts deposited on the skin surface.

When sweat reaches the edge of the pores during the second heating stage, it comes in contact with, wicks into, and spreads over the surrounding salt deposits, rapidly expanding a thin sweat film over the skin. As illustrated in **Figure 8e-f**, this process likely leads to stratum corneum hydration proceeding top-down from a sweat film rather than laterally from the edge of a sweat puddle. Therefore, the spreading of sweat over the skin surface is no longer limited by the lateral stratum corneum hydration rate and induced wetting that confines the liquid into puddles but proceeds faster through salt-deposit facilitated wetting. Since sweat spreads in a thin film manner instead of thicker puddles, a smaller volume is required to cover the surface during the second heating stage. Since the concentration of salts and other molecules in sweat highly vary between individuals [16] and salt deposition patterns depend on both salt concentration and evaporation rate [78], the high variation in the shifts between sweating modes for different subjects observed during second heating stage likely occurs due to variations in both the quantity and patterns of deposited salt. Unfortunately, these deposits are likely very thin and were not detectable using any of the employed imaging techniques.

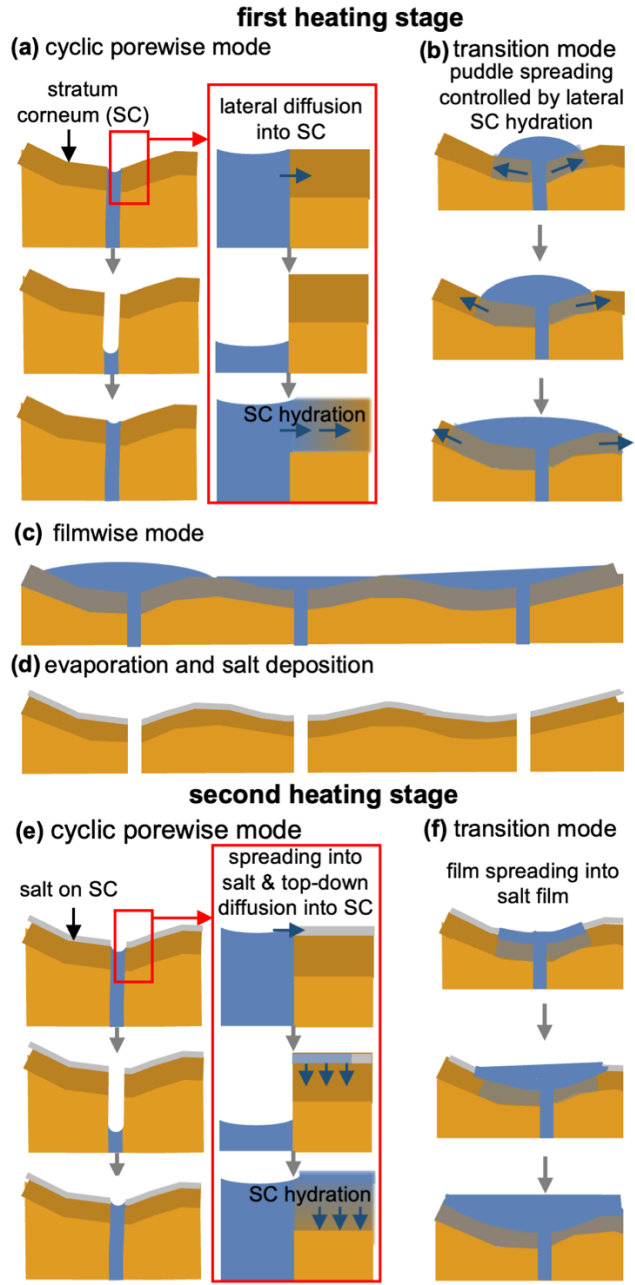

**Figure 8.** Schematics contrasting the difference that the salt depositions induces between sweating during the **(a-c)** first and **(e-f)** second heating stages, that are separted by sweat evaporation and salt deposition **(d)**.

## 5. Conclusions

We conducted an integrative and multi-scale study coupling engineering and physiological methods to explore local human sweat dynamics, combining macroscopic physiological measurements (SR, GSC, and HYD) with advanced microscale surface (MWIR and macrophotography) and cross-sectional (OCT) imaging techniques, providing insights into sweat production, transport, and evaporation mechanisms from insensible perspiration to being "fully soaked" through heating-cooling-heating stages on foreheads of six subjects. In agreement with our prior study employing a mini-wind-tunnel-shaped ventilated capsule [36], after onset, we observed cyclic porewise, transition, and filmwise sweating modes during both heating stages. Using the multi-pronged measurement and imaging approach, we revealed the following sweating characteristics:

1) While the air jet capsule provides much higher evaporation rates, it is also too effective at cooling the skin surface, leading to an artificially depressed sweat rate measurement.

2) In contrast to prior interpretation of GSC and HYD measurements with visual observations [15], using MWIR imaging, we showed that hydration of stratum corneum does not occur uniformly from within the body before sweat emergence on the skin, but instead starts and proceeds in a cyclic and highly heterogenous manner from sweat periodically reaching pores.

3) Our findings challenge the traditional conceptualization of sweat emerging from pores as hemispherical droplets [16,17,36,52,55,69], demonstrating that sweat commonly forms shallow meniscus in the pore. To our knowledge, a high-contact sweat droplet shape has only been shown through OCT by Ohmi et al. [60] for mental sweating on a fingertip. In this location, the stratum corneum is about 25 to 50 times thicker than on the forehead (about 0.5 mm vs. 10 to 20 µm), and ducts and pores have a substantially different shape. In particular, the ducts are spirals within the stratum corneum and end in funnel-shaped pores mostly present on fingertip skin ridges [58,60], rather than in skin crevices and ridges on other body parts [10,11]. We note that we also conducted an exploratory MWIR and OCT imaging of mental sweating on fingertips to determine possible occurrence of hemispherical droplets on the fingertips. However, we could only image relatively shallow and low-contact angle droplets emerging from pores (see the **SM**). In our main results, besides sweat "columns" ending with shallow convex or flat meniscus and flat films, we only observed shallow sweat puddles with contact angles of at most around 40°.

4) In absence of deposited salts, the rate of lateral stratum corneum hydration and induced wetting around the pores and later at the edge of the puddles controls when sweating modes shift in between cyclic porewise (about 15 minutes) to transition and later to filmwise.

5) When deposited salts cover the skin, secondary sweating proceeds more rapidly in a film manner through wicking of the liquid into the salts, bypassing the lateral spreading rate limitations imposed by the rate of lateral hydration stratum corneum from edges of puddles during the first heating stage. Because sweat does not accumulate into thicker puddles on "salty skin", a smaller volume of accumulated sweat is required to cover the skin surface. Since regular skin washing that removes salts is a relatively modern development, the film-only sweating observed during the second heating stage might be considered more "natural". However, the concentration of salts and other molecules in sweat highly varies between individuals [16], likely leading to highly varied salt deposition patterns [78]. Therefore, the salt-facilitated sweat film spreading results in higher variation in the shifts between sweating modes for different subjects.

6) Even on a relatively hair-free forehead, we observed sweat occasionally spreading and evaporating from microscopic hairs, potentially enhancing the sweat evaporation rate.

In the future, our approach could be used to gain further insight into sweating by focusing on other body areas, positions, and activities, skin products (from moisturizer to sunscreen), including a broader set of individuals (e.g., the elderly or with medical conditions know to impact sweating and skin texture) and adding other measurements such as the wind-tunnel shaped ventilated capsules that measure and module the sweat evaporation rate. This comprehensive study substantially advances our understanding of microscale human sweating fundamentals, which can propagate into enhancing applications across clinical diagnostics, textile engineering, and wearable sensor development.

## Acknowledgments

This research was funded by National Science Foundation grant # 2214152 (PIs K. Rykaczewski and S. Kavouras). The authors acknowledge Prof. Heather Emady for providing access to her MWIR camera, Prof. Paul Westerhoff for providing access to his OCT instrument, and Dr. Jnaneshwar Das for granting access to his 3D printer and laser cutter and Devin Keating for his expertise and help using these instruments.

**Data Availability:** Due to large size, the MWIR and OCT movies can be provided upon request.

# Supplemental Material for "A micro-to-macroscale and multi-method investigation of human sweating dynamics"

Cibin T. Jose,[1] Ankit Joshi,[1,2] Shri H. Viswanathan,[1] Sincere K. Nash,[3] Kambiz Sadeghi,[1,2] Stavros A. Kavouras,[4] and Konrad Rykaczewski[1,2*]

1. School for Engineering of Matter, Transport and Energy, Arizona State University, Tempe, USA
2. Julie Ann Wrigley Global Futures Laboratory, Arizona State University, Tempe, AZ 85287, USA
3. School of Life Sciences, Arizona State University, Tempe, AZ 85287
4. College of Health Solutions, Arizona State University, Phoenix, AZ 85004

*Corresponding author: konradr@asu.edu (Orcid ID: 0000-0002-5801-7177)

# S1. The protocol and experimental details
## *S1.1 Detailed human subject information*

**Table S1.1.** Detailed human subject information.

| Subject | Gender | Age (years) | Mass prior to sweating (kg) | Mass after sweating (kg) | Height (cm) |
|---|---|---|---|---|---|
| 1 | Male | 25 | 67.7 | 67.0 | 170.2 |
| 2 | Male | 21 | 94.4 | 93.9 | 182.9 |
| 3 | Female | 21 | 62.7 | 62.4 | 162.6 |
| 4 | Male | 21 | 86.0 | 85.6 | 185.4 |
| 5 | Female | 20 | 74.6 | 74.2 | 167.3 |
| 6 | Female | 19 | 59.0 | 58.5 | 167.6 |
| 7 | Male | 41 | 70.3 | 69.7 | 175 |

## *S1.2 Study participant details*

The Arizona State University Institutional Review Board approved the human subject experiment. All participants were screened using inclusion criteria (between 18 and 55 years old with no history of significant health issues such as high blood pressure or current symptoms that mild hyperthermia might exacerbate) and provided written informed consent before participating. The subjects were asked not to drink alcohol the evening prior to the experiment, have a light breakfast without caffeine, wash their face only with water, not apply any skin care products, and hydrate with at least 0.5 L of water two hours before the experiment, which typically started around 8:30-9:00 in the morning and lasted about 2.5 hours. The experiments were conducted between February 14th and 26th of 2025 in Walton Center for Planetary Health on Arizona State University, Tempe, Arizona campus. The laboratory temperature, relative humidity, and air speed (measured using Campbell Scientific Windsonic4 2D anemometer) were 20 to 22°C, relative humidity 30 to 40%, and below 0.1 m×s$^{-1}$, respectively. We conducted experiments with six healthy, non-heat acclimated subjects: three females (age: 22.3±2.3 years, weight: 82.6±13.7 kg, height: 180.4±8.8 cm) and three males (age: 20±1 years, weight: 65.4±8.2 kg, height: 165.3±2.9 cm) with Caucasian, Indian, and African American decent. We deemed this number of subjects sufficient since it was an observational study looking at representative sweating dynamics on a single skin site, and no significant gender differences were observed in a related study by Gerrett et al.[1]. In addition, we also conducted a pilot comparing the air jet, 1.7 cm$^2$ cylindrical, and 2.7 cm$^2$ cylindrical (matching exposed skin area of the jet capsule) ventilated capsules on the forehead of a 41 years old male, had a height of 175 cm, and weighed 69 kg.

Upon arrival, participants were asked to weigh themselves without any clothes, and this measurement was repeated after the experiment to calculate total sweat loss. The subjects wore a long-sleeved shirt and pants underneath the liquid-perfused full-body suit. Before each experiment, the capsule was purged with dry air to eliminate any moisture absorbed from the surrounding laboratory air. Since our prior study found that only a mild core temperature increase occurred during the heating process (0.5°C and 0.8°C), we did not monitor this value in current experiments. After the subject laid down, the capsule was positioned on their forehead, and GSC electrodes were attached to their forehead, leaving room for the periodic (2-3 minute) HYD measurement. The subject instrumentation process took about 15 to 20 minutes, during which baseline values of parameters, including imaging, were taken. Afterwards, the subjects were heated with water at 47°C and the electric blankets. As assessed by GSC and MWIR, sweating onset typically occurred after about 15 to 25 minutes of heating and was continued until a sweat layer entirely covered the forehead. Subsequently, the subjects were cooled using 25°C water with the exterior blanket removed until SR and HYD measurements returned to the pre-heating baseline (GSC was always higher due to film under electrodes) and re-heated using the same procedure until their foreheads were again flooded by sweat. The subject's comfort level was checked verbally throughout, with the experiment ending if discomfort occurred. Subject 6 did not feel comfortable after the first heating stage, so we did not proceed with the second heating stage in this case. If requested, the subjects were provided with a measured amount of water throughout the experiment. At the end of the experiment, the participant was asked to weigh themselves again, rest, and hydrate for 15 minutes. A net water loss of around 0.4 to 0.7 kg was noted, accounting for weight and water intake changes. We note that we also took representative water droplet (~1 mL of distilled water dispensed using micropipette) images near the imaged area before the first sweating cycle and after both the first and second sweating stages using a Sony A7Siii camera with Venus Optics Laowa 24mm f/14 Probe Lens. The water contact angles before heating varied between 30 to 75° on dry skin and were below 10° during sweating, as in our previous work [2]. However, as is evident in our microscopic MWIR and OCT images, these contact angles are not representative of the bare wetting properties of the skin, as microscale hairs and surface topography impact droplets. Consequently, the contact angles are not meaningful measures of the wetting properties and are not utilized.

### S1.3 Dielectric epidermal hydration (HYD) meter and Galvanic Skin Conductance (GSC) measurements

The HYD was measured with 2-to-3-minute frequency using a capacitive skin hydration meter (Delfin MoistureMeter SC, Delfin Technologies, Kuopio, Finland) in an area that was 2 to 4 cm from the imaged center of the forehead (see **Figure 1b** in the main manuscript). According to the manufacturer, the instrument has an operating frequency of 1.3 MHz and resolution of 0.1% and provides output in arbitrary units (AU) between 0 and ~180 that reflect the combined effects of capacitance and dielectric constant of the outer ~50 mm of skin [3]. The thick pen-shaped meter has an internal force meter that indicates the appropriate force to apply for the measurement (about 1.5 to 2 N [1]), which takes about 5 to 10 seconds to complete. The baseline HYD reading is typically between 20 and 60 AU and increases gradually to 100 and beyond when the skin is covered by sweat.

We have measured the GSC using the MP36 instrument (BSL Basic System, BSLBSC-W, Biopac). Two EDA electrodes (EL507A, Biopac) are connected to the forehead at least 5 cm from each other. Before attaching the electrodes, an isotonic Electrode gel for EDA (GEL101A, Biopac) from BIOPAC Systems, Inc. is applied under the electrode. The system measures the resulting current, which is used to compute skin conductance and is measured in microsiemens (μS) with an instrument range of 0.1 to 100 μS. The electrodes (EL507A, Biopac) are connected to the CH3 via SS57LA transducer for data acquisition. The SS57LA electrodermal activity transducer operates by applying a constant voltage of approximately 0.5 V across the electrodes and then measures the current flowing between the two electrodes. Because the voltage (V) is fixed, from Ohms Law, the conductance (G) will be proportional to the current (I); therefore, G = I/V = I/0.5 V. The software then performs the necessary scaling and unit conversion, and the preset Electrodermal Activity (EDA, SS57LA, Biopac), 0-35 Hz is selected, and the acquisition rate of 2,000 kHz is chosen during the experiment. Later, during post-processing, the sample is reformatted to 0.2 Hz, corresponding to one data point every 5 seconds. The baseline conductivity of the skin prior to any sweating varied between 3 and 10 mS, which was comparable to the increase in the conductance with the first sweating cycle. To facilitate interpretation of the data, we follow Gerrett et al. [1] and report the values as the difference above the baseline (DGSC).

### S1.4 Midwave Infrared imaging

We captured the MWIR videos using a FLIR MWIR 6701 camera (with a 50 mm f2.5 macro lens) with detector with a 3 to 5 μm spectral range and 640 × 512 pixel count [2]. We recorded the images at 10 Hz with the lens zoom extended to provide an image of the central forehead skin area (16.4 mm by 11.5 mm typically) with scale obtained by imaging a reference metal ruler (typically translating to about 25 μm per pixel spatial resolution). We exported the captured movies (.ats format) from the FLIR Research Studio software into .avi format and analyzed them using FIJI/ImageJ 2.14.0.

### S1.5 Optical Coherence Tomography imaging

We conducted OCT imaging using the Ganymade-II OCT system from Thorlabs, which operates on the near-infrared (NIR) range with a center wavelength of 930nm. The instrument has a scan (pixel to pixel) rate of up to 248 kHz, imaging depth of up to 3.4 mm, and axial resolution in the range of 3 to 6 μm. The OCT system is equipped with a rigid scanner for high stability and ease of use, which is then mounted on an arm for mobility. The areas of interest (e.g., sweat droplets) were identified using the "macrophotography" camera integrated into the OCT system, images from which we also include. We found that the skin's movement and microscopic vibration required the acquisition of two-dimensional cross-sectional (B-scan) images in a "movie" mode with about 400 images per sequence. We fixed the imaging depth at 2.2 mm and varied the scan length on the surface depending on the desired cross-section. The system automatically adjusted the per image acquisition rate based on the field of view (FOV) length, with the overall range being 14 to 35 Hz. We exported the captured image sequence (.oct format) from the ThorImage software into .tiff format and analyzed them using FIJI/ImageJ 2.14.0.

## S2. Cylindrical and air jet ventilated capsule design details

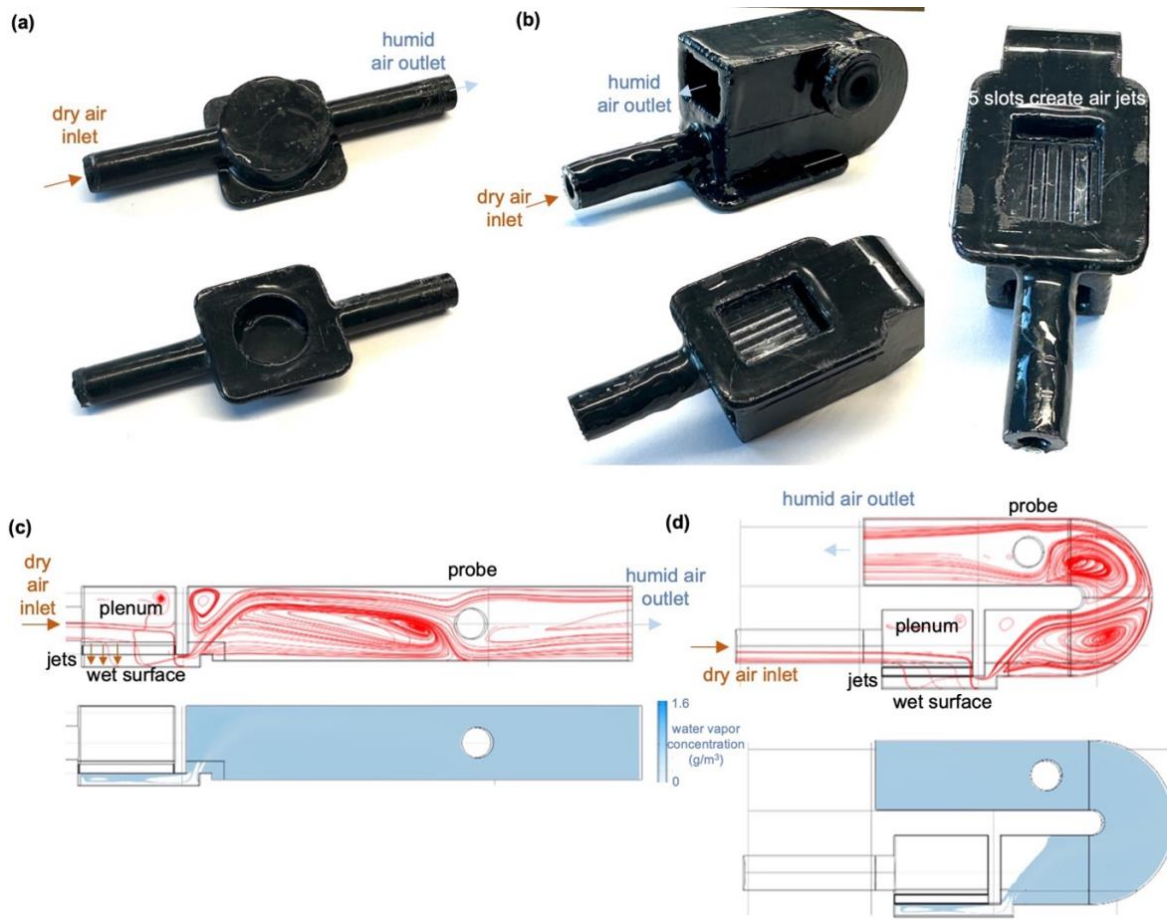

**Figure S2.1** Photographs of **(a)** the 2.7 cm2 cylindrical capsule and **(b)** the U-bend air jet capsule and simulated flow streamlines and water concentration fields within a **(c)** straight and **(d)** U-bend air jet capsules. The simulation setup is described in Jaiswal et al. [2] and Rykaczewski [4].

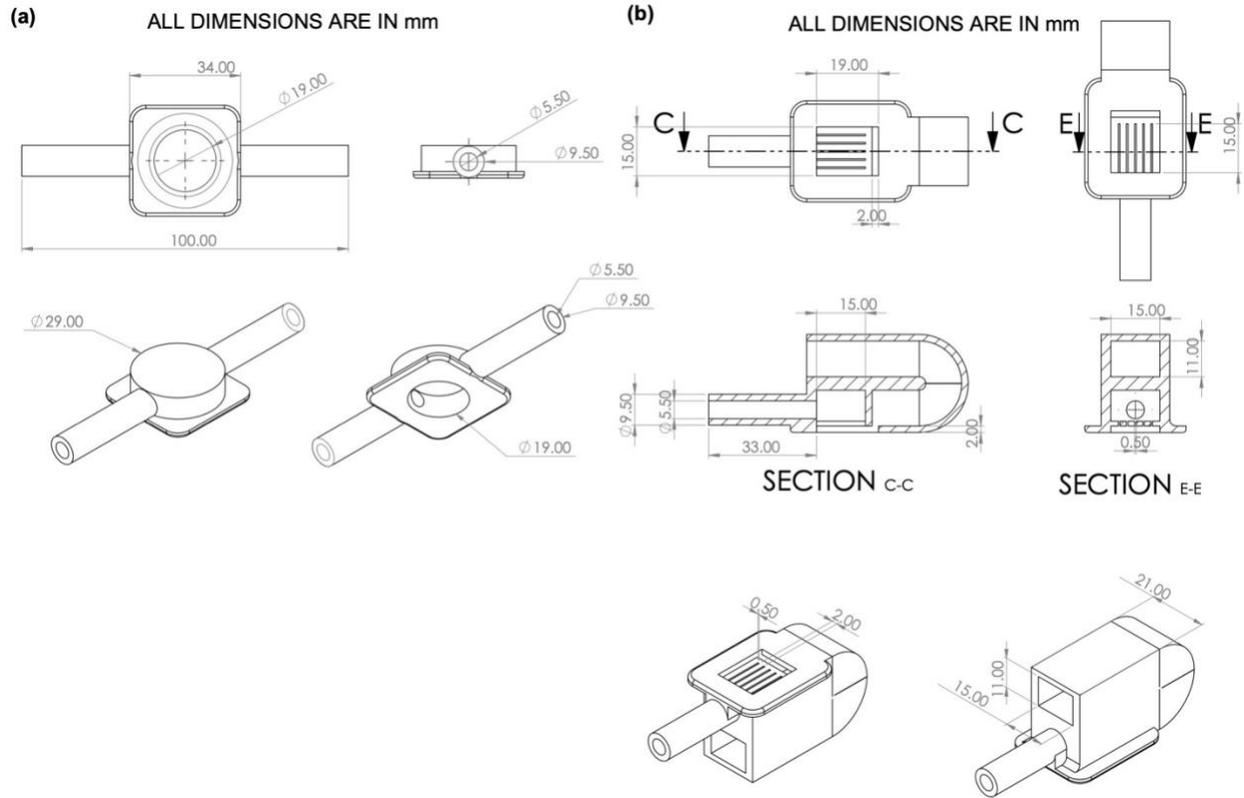

**Figure S2.** Mechanical drawings of **(a)** the 2.7 cm$^2$ cylindrical capsule and **(b)** the 2.7 cm$^2$ U-bend air jet capsule.

**S3. The complete SR, HYD, GSC, MWIR, macrophotography, and OCT results**

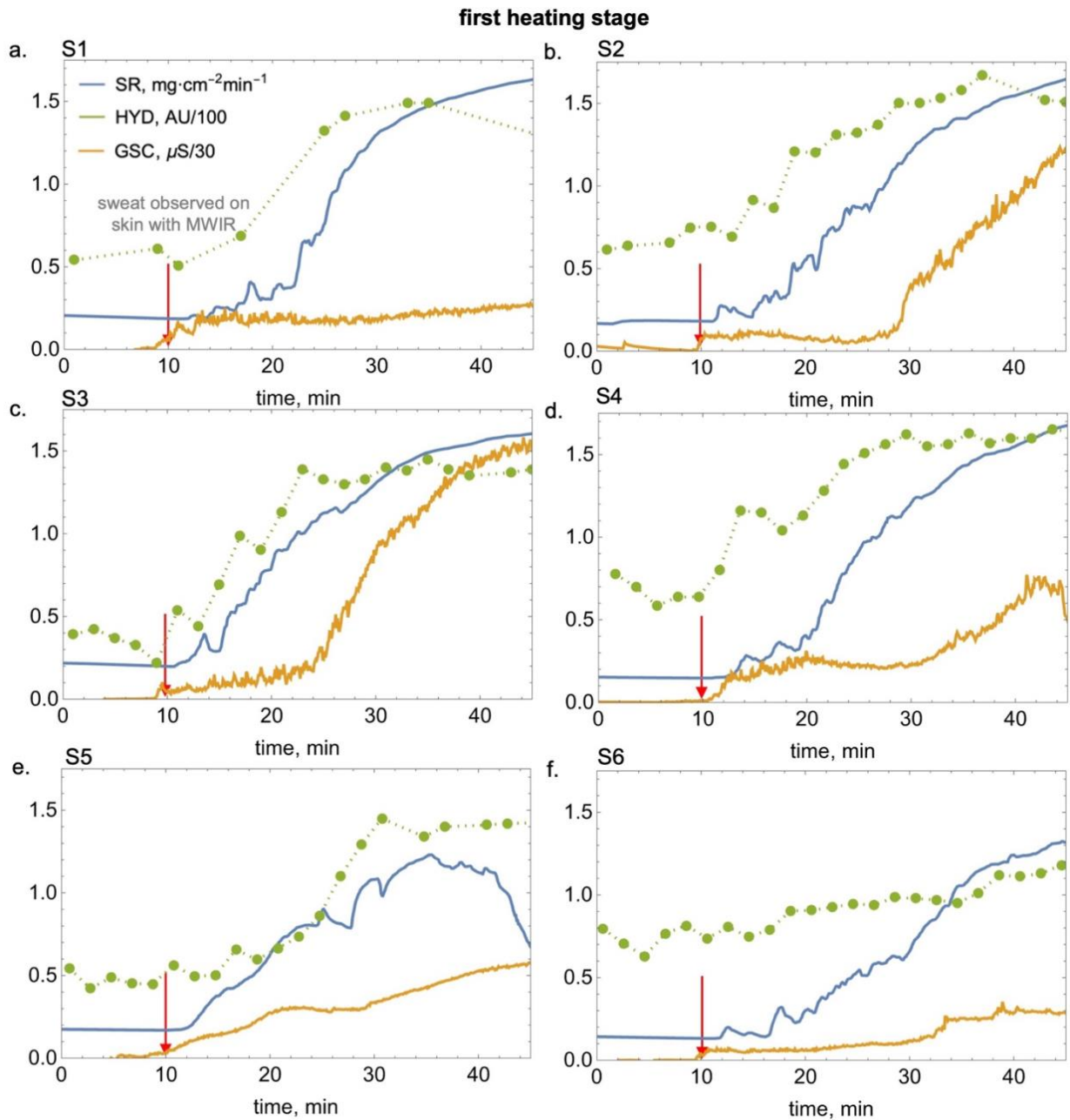

**Figure S3.1** The sweat rate (SR), galvanic skin conductance (ΔGSC, from baseline), and dielectric epidermis hydration (HYD) plotted against time for the first heating stage for subjects 1-6 (note: the ΔGSC and HYD measurements are scaled as indicated in legend to facilitate comparison). The timeframe is adjusted so that onset of sweat detected with MWIR is at 10 minutes.

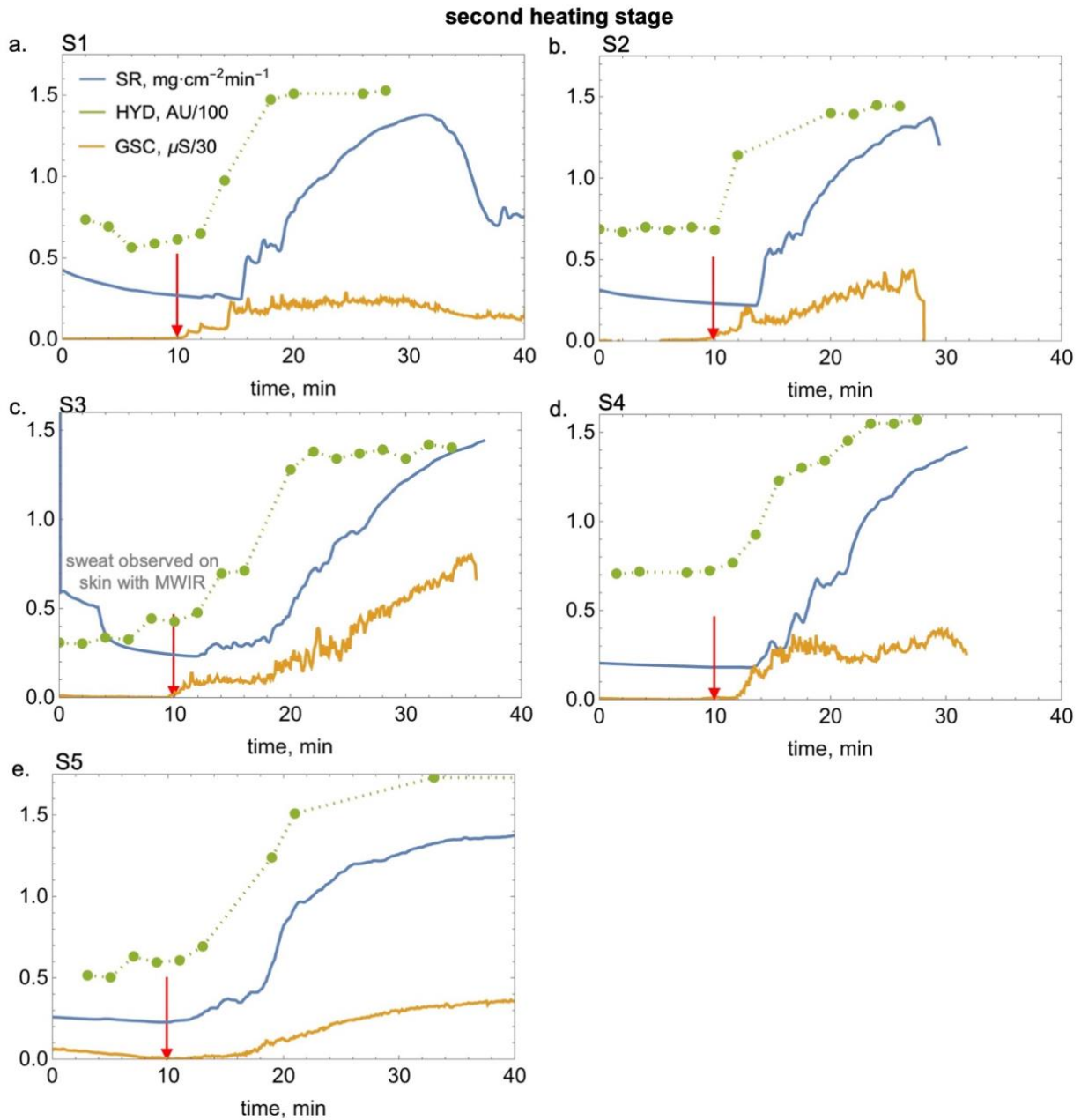

**Figure S3.2** The sweat rate (SR), galvanic skin conductance (ΔGSC, from baseline), and dielectric epidermis hydration (HYD) plotted against time for the second heating stage for subjects 1-5 (note: the ΔGSC and HYD measurements are scaled as indicated in legend to facilitate comparison and subject 6 did not complete this stage). The timeframe is adjusted so that onset of sweat detected with MWIR is at 10 minutes.

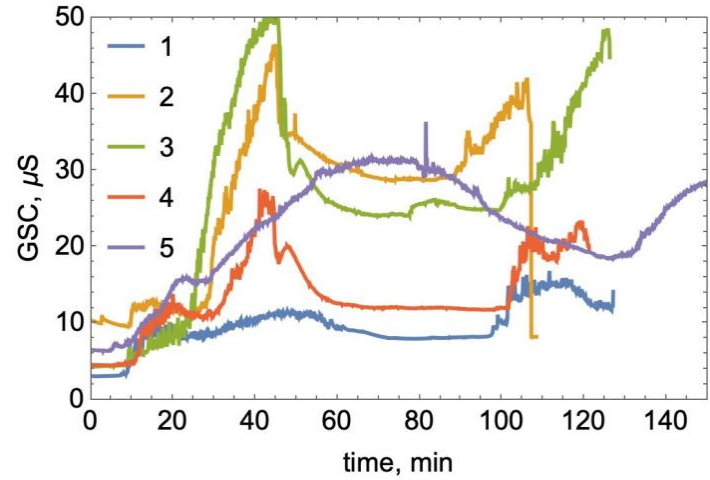

**Figure S3.3** The absolute galvanic skin conductance (GSC) plotted against time for subjects 1-5 (subject 6 did not complete the second heating stage due to discomfort). The timeframe is adjusted so that onset of sweat detected with MWIR is at 10 minutes.

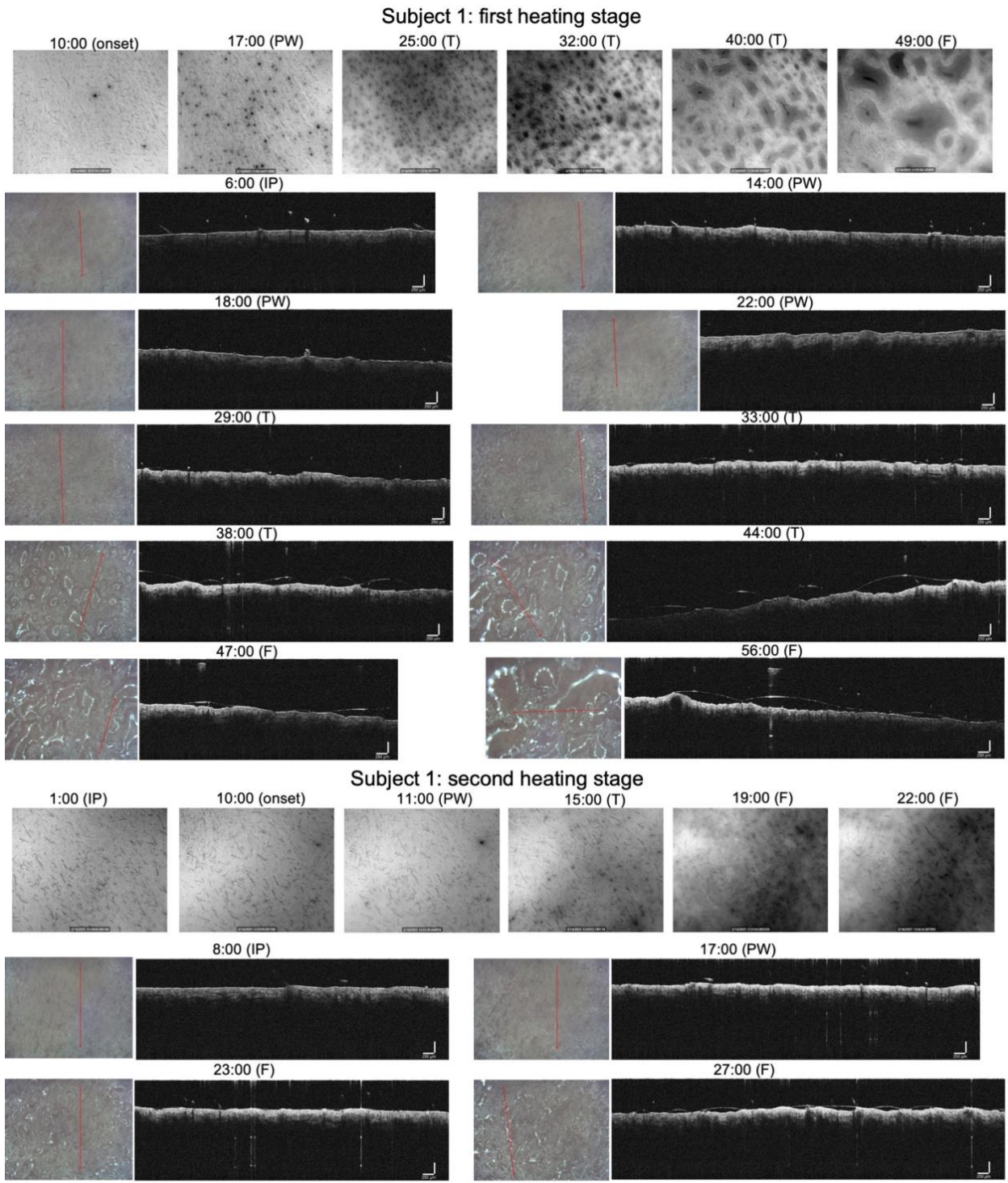

**Figure S3.4** The MWIR, macrophotography, and OCT images for Subject 1 (the redline in photograph indicates the OCT scan). The timeframe is adjusted so that onset of sweat detected with MWIR is at 10 minutes.

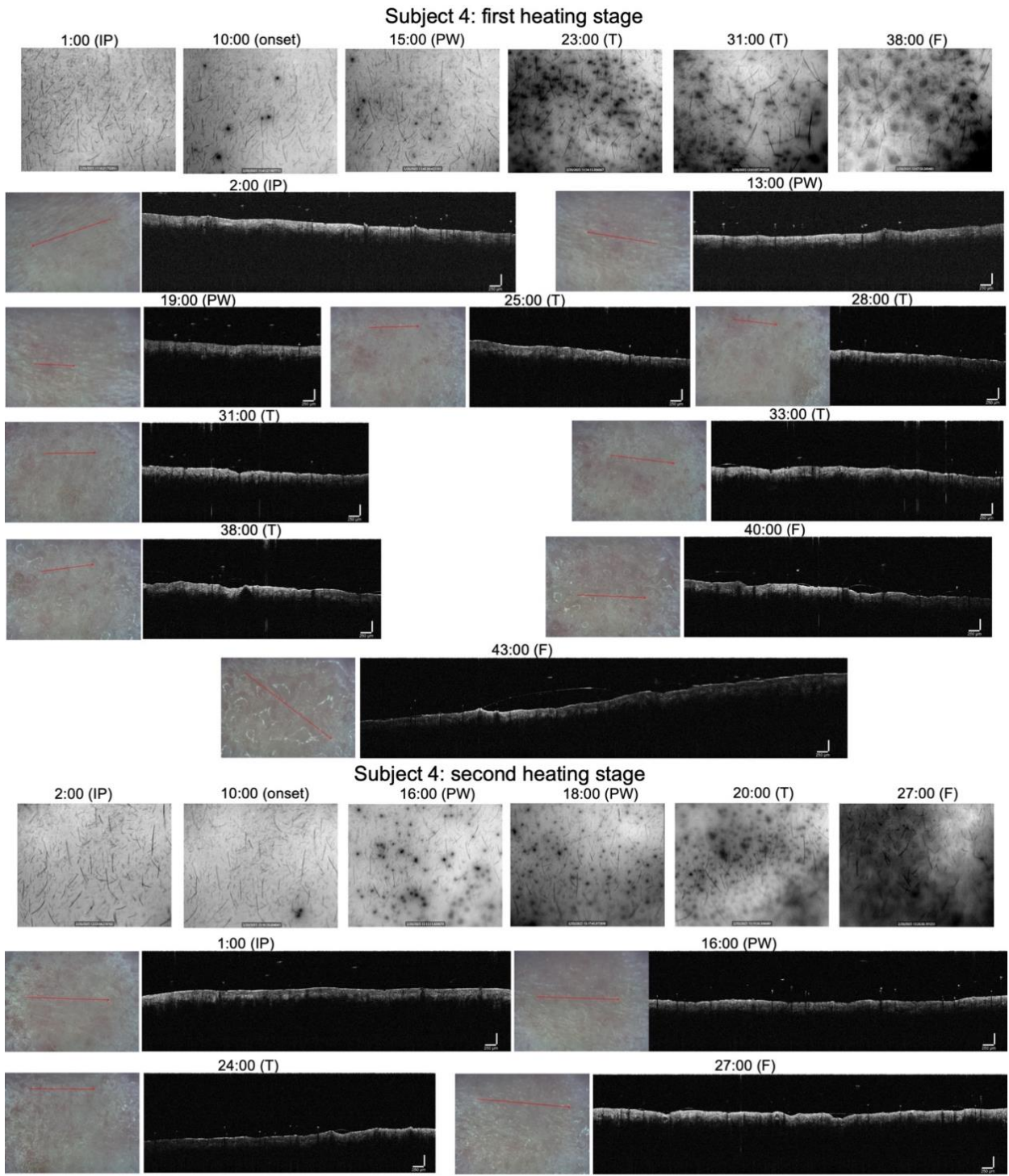

**Figure S3.5** The MWIR, macrophotography, and OCT images for Subject 4 (the redline in photograph indicates the OCT scan). The timeframe is adjusted so that onset of sweat detected with MWIR is at 10 minutes.

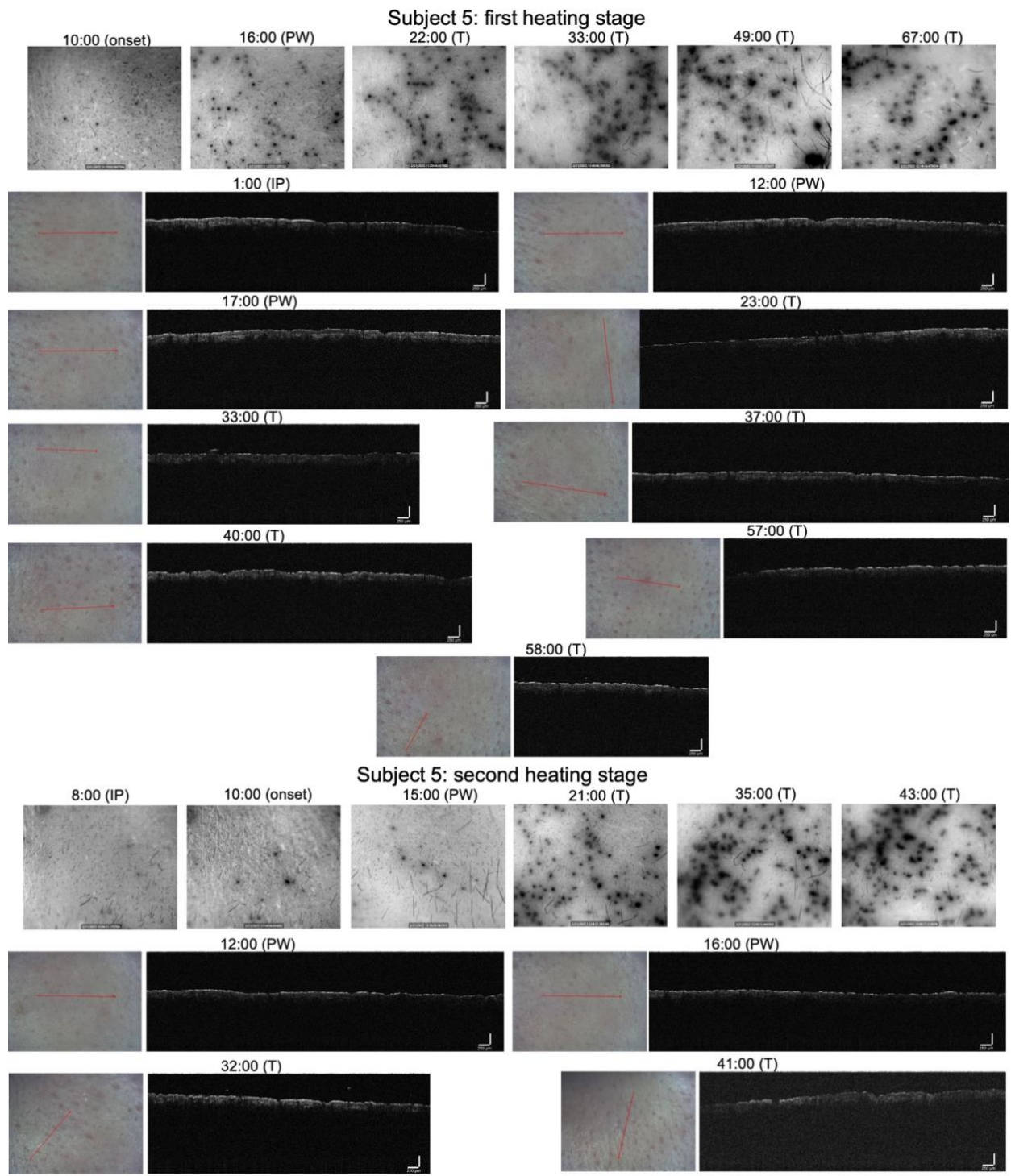

**Figure S3.6** The MWIR, macrophotography, and OCT images for Subject 5 (the redline in photograph indicates the OCT scan). The timeframe is adjusted so that onset of sweat detected with MWIR is at 10 minutes.

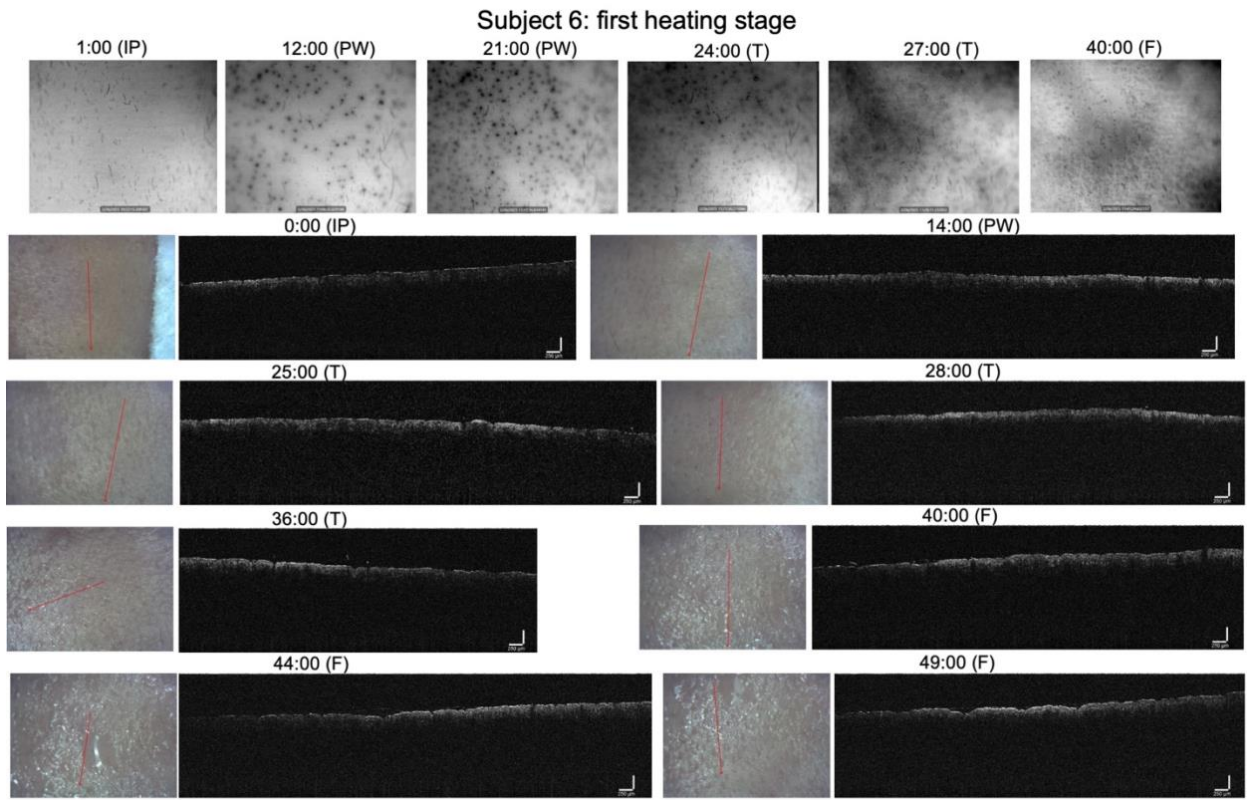

**Figure S3.7** The MWIR, macrophotography, and OCT images for Subject 6 (the redline in photograph indicates the OCT scan). The timeframe is adjusted so that onset of sweat detected with MWIR is at 10 minutes.

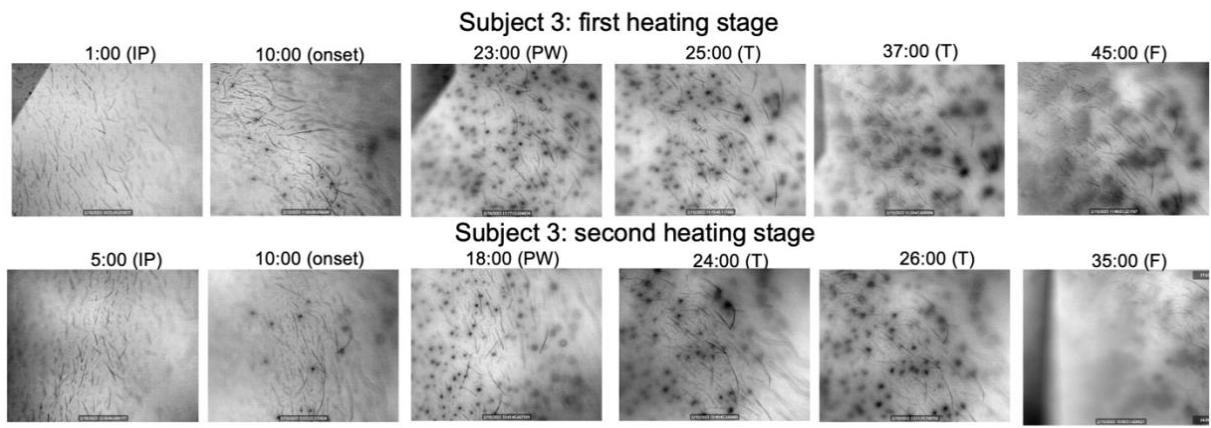

**Figure S3.8** The MWIR for Subject 3 (the OCT malfunctioned during this experiment). The timeframe is adjusted so that onset of sweat detected with MWIR is at 10 minutes.

## S4. Representative sweat drying dynamics

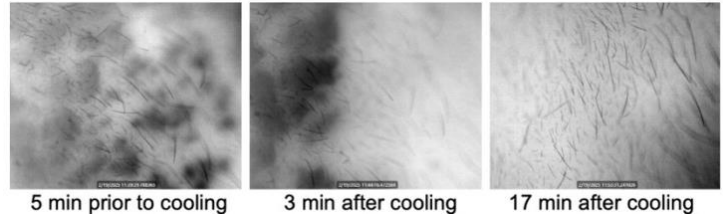

5 min prior to cooling    3 min after cooling    17 min after cooling

**Figure S4.1** The MWIR for Subject 3 showing that sweat evaporates from the skin within about 15 minutes after the start of subject cooling.

## S5. Representative sweat on fingertip

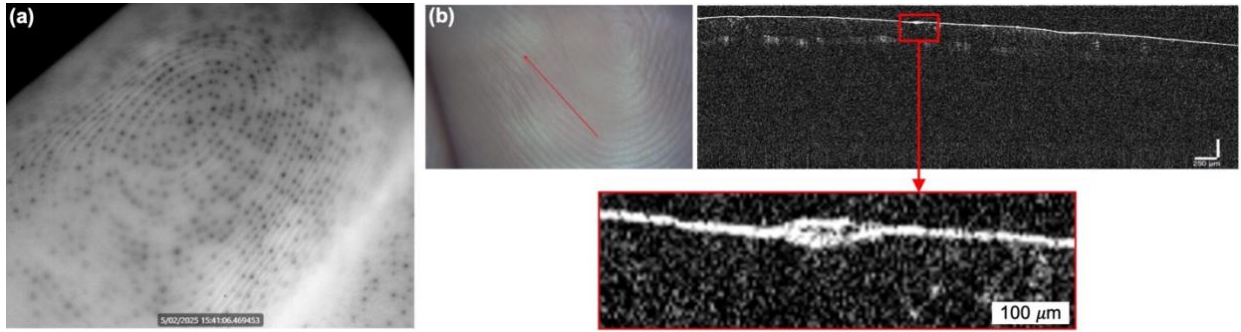

**Figure S5.1 (a)** MWIR and **(b)** macrophotography and OCT of sweat emerging from pores on a fingertip.